\documentclass[superscriptaddress,nofootinbib,twocolumn,letterpaper,prx]{revtex4-1}

\usepackage{amsmath,amssymb}
\usepackage{graphicx}
\usepackage[usenames,dvipsnames]{xcolor}
\usepackage{hyperref}
\usepackage{url}
\usepackage[utf8]{inputenc}
\usepackage{slashed}
\usepackage{subfigure}
\usepackage{slashed,bbm}
\usepackage{graphicx,psfrag,epsfig}
\usepackage{dsfont}
\usepackage{setspace}
\usepackage{blindtext}
\hypersetup{
    colorlinks,
    linkcolor={red},
    citecolor={blue},
    urlcolor={blue!40!black}
}

\allowdisplaybreaks

\begin{document}

\title{Critical behavior of Dirac fermions from perturbative renormalization}

\author{Bernhard Ihrig}
\affiliation{Institute for Theoretical Physics, University of Cologne, Z\"ulpicher Stra\ss e 77, 50937 Cologne, Germany}

\author{Luminita N. Mihaila}
\affiliation{Institute for Theoretical Physics, Heidelberg University, Philosophenweg 16, 69120 Heidelberg, Germany}

\author{Michael~M.~Scherer}
\affiliation{Institute for Theoretical Physics, University of Cologne, Z\"ulpicher Stra\ss e 77, 50937 Cologne, Germany}

\begin{abstract}
Gapless Dirac fermions appear as quasiparticle excitations in various condensed-matter systems.
They feature quantum critical points with critical behavior in the 2+1 dimensional Gross-Neveu universality class.
The precise determination of their critical exponents defines a prime benchmark for complementary theoretical approaches, such as lattice simulations, the renormalization group and the conformal bootstrap.
Despite promising recent developments in each of these methods, however, no satisfactory consensus on the fermionic critical exponents has been achieved, so far.
Here, we perform a comprehensive analysis of the Ising Gross-Neveu universality classes
based on the recently achieved four-loop perturbative calculations.
We combine the perturbative series in  $4-\epsilon$ spacetime
dimensions with the one for the purely fermionic Gross-Neveu model in $2+\epsilon$ dimensions
by employing polynomial interpolation as well as two-sided Pad\'e approximants.
Further, we provide predictions for the critical exponents exploring various resummation techniques following the strategies developed for the three-dimensional scalar $O(n)$
universality classes.
We give an exhaustive appraisal of the current situation of Gross-Neveu universality by comparison to other methods.
For large enough number of spinor components $N\geq 8$ as well as for the case of emergent supersymmetry $N=1$, we find our renormalization group estimates to be in excellent agreement with the conformal bootstrap, building a strong case for the validity of these values.
For intermediate $N$ as well as in comparison with recent Monte Carlo results, deviations are found and critically discussed.
\end{abstract}

\maketitle

\section{Introduction}

Gapless Dirac fermions with quasirelativistic dispersion emerge as low-energy excitations in many different condensed-matter systems, for example in graphene, $d$-wave superconductors and surface states of topological insulators~\cite{doi:10.1146/annurev-conmatphys-031113-133841,doi:10.1080/00018732.2014.927109}.
In particular, in two spatial dimensions, interacting Dirac fermions feature quantum critical points~\cite{0295-5075-19-8-007,PhysRevLett.87.246802,PhysRevLett.97.146401,PhysRevLett.100.146404,PhysRevLett.100.156401,PhysRevB.79.085116}.
A representative example is given by the semimetal-to-insulator transition of interacting electrons on graphene's honeycomb lattice, where the ordered state corresponds to a sublattice symmetry broken insulating state with charge order~\cite{PhysRevLett.87.246802,PhysRevLett.97.146401,Chandrasekharan:2013aya}.
Other examples for continuous transitions of Dirac electrons can be constructed on suitable lattice models, e.g., with magnetic, superconducting and more exotic transitions~\cite{PhysRevB.91.165108,PhysRevX.6.011029,2015arXiv151207908L,PhysRevB.94.205136,PhysRevB.96.115132,PhysRevB.97.125137,Li:2017dkj,PhysRevB.96.195162,Otsuka:2018kcb,2018arXiv180500478X,PhysRevB.95.241103}.
Near these quantum critical points, universal critical behavior occurs as expressed in a collection of simple power laws for physical quantities, such as the order parameter, together with their critical exponents which characterize the universality class~\cite{ZinnJustin:2002ru,herbut_2007,sachdev_2011}.

Generally, the quantitatively precise determination of critical exponents at any continuous phase transition and, hence, the identification of the corresponding universality class defines a benchmark for complementary theoretical methods.
A pivotal role in this context is played by the universality classes of the three-dimensional scalar $O(n)$~models, which represent the critical behavior for a broad range of continuous phase transitions from different areas of physics.
To describe these universality classes various theoretical methods have been applied, which culminated in an impressive agreement across the different approaches ranging from field theoretical renormalization group (RG) studies~\cite{0305-4470-31-40-006,Kompaniets2017} over Monte-Carlo simulations~\cite{PhysRevB.74.144506,PhysRevB.82.174433,PhysRevB.84.125136} to the conformal bootstrap~\cite{Kos:2016ysd}.
While the quest for precision calculations of the $O(n)$ universality classes has been going on for decades, it was only very recently that the conformal bootstrap program produced estimates for the critical exponents with an accuracy well beyond the corresponding Monte Carlo simulations~\cite{Kos:2016ysd}.
Motivated by this development and due to additional insights into the mathematical structure of Feyman diagrams, also the RG was recently pushed to six-loop order for the scalar $O(n)$ models~\cite{Kompaniets2017,Batkovich:2016jus,Kompaniets:2016hct}.
Together with appropriate resummation of the resulting asymptotic series, the RG provides very accurate results for the critical exponents in three dimensions, which are in excellent agreement with the other methods~\cite{Kompaniets2017}.
Efforts to push the boundary to even higher precision are ongoing~\cite{PhysRevD.97.085018}, but this line of research can already be considered as extremely successful as it is unparalleled to achieve such convincing quantitative agreement for highly non-trivial physical quantities in statistical physics across different approaches.

Critical gapless Dirac fermions in 2+1 dimensions are not captured by the Ising or $O(n)$ universality classes.
Instead, they can effectively be described by chiral transitions appearing in different variants of the three-dimensional Gross-Neveu or Nambu-Jona-Lasinio models -- defining the chiral universality classes~\cite{Rosenstein:1993zf,PhysRevLett.97.146401,Fei2016,PhysRevB.97.041117}.
The Gross-Neveu~\cite{PhysRevD.10.3235} and Nambu-Jona-Lasinio~\cite{PhysRev.122.345} models incorporate a high-degree of symmetry, which is emergent in the described physical systems at low energies, including relativistic or chiral invariance~\cite{PhysRevB.79.085116}.
They avoid many complications which are typically present in most interacting fermion models with lower degree of symmetry, e.g., due to a finite Fermi surface. Therefore, they are better accessible by theoretical approaches, i.e. they define a clean starting point for the study of critical behavior in interacting many-fermion models.

In view of the successful description of the critical behavior of the three-dimensional $O(n)$ models, it is tempting to believe that a similar precision can also be achieved for the case of the comparatively simple three-dimensional Gross-Neveu models.
Indeed, there has been promising progress in the development of the various methods, recently, suggesting that the consensual precision determination of the Gross-Neveu universality classes is within reach:
(1)~Numerical approaches have found sign-problem free formulations for the calculation of various important quantum phase transitions of interacting Dirac fermions on the lattice~\cite{Chandrasekharan:2013aya,1367-2630-16-10-103008,1367-2630-17-8-085003,PhysRevD.96.114502,PhysRevB.93.155157,PhysRevX.6.011029,Li:2017dkj,2015arXiv151207908L,Otsuka:2018kcb,2018arXiv180508219L}.
(2)~The conformal bootstrap has emerged as a numerical tool to determine critical exponents for fermionic models~\cite{Bashkirov:2013vya,Bobev:2015vsa,Iliesiu:2015qra,Iliesiu:2017nrv}.
(3)~Nonperturbative field-theoretical methods like the functional renormalization group (FRG) have managed to explore sophisticated truncation schemes~\cite{Janssen:2014gea,Vacca2015,Gies2017,Knorr:2017yze,PhysRevB.94.245102,Feldmann:2017ooy}.
(4)~The perturbative renormalization group (pRG) has seen substantial advances in computational technology and the development of suitable algorithms which facilitate the calculation of higher-loop orders.
By now, the pRG has been employed for up to four-loop calculations for Gross-Neveu and similar models in Refs.~\cite{Gracey:2016mio,PhysRevD.96.096010,Gracey:2018qba}.
Despite these promising developments, however, no satisfactory agreement across different theoretical methods has been found for the fermionic universality classes, yet.

In this work, we perform a thorough analysis of resummation and interpolation techniques within the perturbative renormalization group approach to extract quantitative renormalization group predictions for the critical exponents.
We show that this strategy reconciles discrepancies between the conformal bootstrap results from Ref.~\cite{Iliesiu:2017nrv} and the pRG calculations, but not with Quantum Monte Carlo simulations.
To that end, we focus on the simplest version of the Gross-Neveu-Yukawa models, i.e. the chiral Ising model~\cite{Rosenstein:1993zf}, which in $2<D<4$ lies in the same universality class as the purely fermionic Gross-Neveu model~\cite{HASENFRATZ199179,ZINNJUSTIN1991105}.
Even this simplest model has a number of interesting applications.
Most prominently, for $N=8$, the theory describes the quantum critical point of the semimetal-to-insulator transition of spin-$1/2$ electrons on the graphene lattice. In the insulating phase  sublattice symmetry is broken and charge density wave (CDW) order occurs~\cite{PhysRevLett.97.146401}.
The eight spinor components originate from the two sublattices of the honeycomb lattice, the two inequivalent Dirac points in the Brillouin zone and two spin projections of the spin-$1/2$ electrons.

Another application of the Gross-Neveu model is the case of $N = 4$.
According to the counting of spinor components in the graphene case, this corresponds to a model of spinless fermions on the honeycomb lattice.
Strong repulsive nearest-neighbor interactions also drive the spinless system through a semimetal-to-insulator transition~\cite{PhysRevLett.100.156401}.
This simplified version of graphene is accessible to a broad range of different numerical methods with reduced computational cost and therefore has been extensively studied, previously~\cite{1367-2630-16-10-103008,1367-2630-17-8-085003,PhysRevD.96.114502,PhysRevB.93.155157,Vacca2015,PhysRevB.94.245102,Fei2016,Iliesiu:2017nrv,Mihaila:2017ble,PhysRevD.96.096010}.
For $N = 1$, it has been argued that in $D=3$ a minimal $\mathcal{N}=1$ superconformal theory emerges from the Gross-Neveu-Yukawa model at the quantum
critical point which might be relevant at the boundary of a topological phase~\cite{Grover280,Fei2016}.
In the following, we study these particular cases in depth and provide improved estimates for their critical exponents. We carefully compare with other methods.

\subsection{Outline}

In the next section, we first introduce the Gross-Neveu model, Sec.~\ref{sec:GNmodel},  and the Gross-Neveu-Yukawa model, Sec.~\ref{sec:GNYmodel}.
We then shortly summarize the renormalization group procedure in Sec.~\ref{sec:RGmethod}. For more details, we would like to refer the reader to our recent publication, cf.~Ref.~\cite{PhysRevD.96.096010}.
Further, we introduce the notation for the renormalization group functions in Sec.~\ref{sec:RGfunctions}, where we also recall the fixed-point calculation and the relation to the critical exponents, i.e. the correlation length exponent and the anomalous dimensions.
Then, in Sec.~\ref{sec:Pade}, we systematically study the Pad\'e approximants of the perturbation series for the critical exponents of the Gross-Neveu-Yukawa model, i.e. the inverse correlation length exponent and the boson and fermion anomalous dimensions. Therefore, we first calculate the full range of estimates from the Pad\'e approximants in 2+1 dimensions as extracted from the series in $4-\epsilon$ dimensions.
Further, we append a study of the two-sided Pad\'e approximants by combining the $4-\epsilon$ series with the one for the Gross-Neveu model in $2+\epsilon$ dimensions.
Sec.~\ref{sec:polyinterpolation} also explores an interpolation between $2+\epsilon$ and $4-\epsilon$ dimensions, however, with a simpler polynomial expansion and exhibits convergence with the two-sided Pad\'e approximants upon increasing the order of the underlying power series.
In Sec.~\ref{sec:Borelresummation}, we apply the more sophisticated Borel resummation techniques, which were successfully employed in the scalar $O(n)$ models and study their stability with respect to the various resummation parameters.
We discuss specific applications to graphene, spinless fermions on the honeycomb lattice, emergent supersymmetry and the large-$N$ situation.
We conclude in Sec.~\ref{sec:conculsions} by discussing the possible origin of the remaining discrepancies between the different theoretical approaches and add suggestions for future studies.
Further technicalities are given in the appendix.

\section{Models and method}\label{sec:models}

There are two different models which can be used to provide estimates for the critical behavior of 2+1 dimensional Dirac fermions, namely the Gross-Neveu (GN) model and the Gross-Neveu-Yukawa (GNY) model.
Both models exhibit a continuous chiral phase transitions of interacting Dirac fermions but with different dimensionality, offering the opportunity to approach the physically interesting case of $D=3$ dimensions from two sides.
We will shortly introduce the two models in this section.
Further, we summarize the main aspects of the applied renormalization group procedure to obtain the full four-loop order perturbative series.
Since our focus lies on the determination of critical exponents from previously derived renormalization group functions, we keep this discussion short and refer to Refs.~\cite{Gracey:2016mio,PhysRevD.96.096010} for more details.
%

\subsection{Gross-Neveu model}\label{sec:GNmodel}

The Gross-Neveu (GN) model~\cite{PhysRevD.10.3235} is a simple purely fermionic quantum field theory of spin-1/2  Dirac fermions which
interact via a four-fermion interaction.
The Lagrangian of the GN model reads
\begin{align}
  \mathcal{L}_\text{GN} =  \bar{\psi}_i \slashed{\partial} \psi_i  + \frac{1}{2} g (\bar{\psi}_i \psi_i)^2 \,, \label{eq:GNLagrangian}
\end{align}
and we work in $D$-dimensional euclidean spacetime.
The field $\psi = (\psi_i)$ with $i\in \{1,\dots,N_\text{f}\}$ denotes a $N= d_\gamma N_\text{f}$ component spinor. Here, $N_\text{f}$ is the number of fermion flavors with $d_\gamma=4$ being the dimension of the Clifford algebra of a single flavor of four-component Dirac fermions.
We use the notation $\slashed{\partial}=\gamma_\mu\partial_\mu$ with $\mu, \nu \in\{0,1,...D-1\}$ and the conjugate of the Dirac field is given by $\bar\psi=\psi^\dagger\gamma_0$.
The coupling constant $g$ is dimensionless in $D=2$ where the model is perturbatively renormalizable.
For $D>2$, a large-$N$ expansion can be performed~\cite{Gracey:1990wi,Gracey:1992cp,Vasiliev:1992wr,Vasiliev:1993pi,Gracey:1993kb,Gracey:1993kb,Gracey:1993kc,Derkachov:1993uw,Gracey:2017fzu,Manashov:2017rrx} and also expansion around the lower critical dimension, i.e. $D=2+\epsilon$ expansion, has been studied, see, e.g., Refs.~\cite{Gracey:1990sx,Gracey:1991vy,Luperini:1991sv,Janssen:2014gea,Fei2016,Gracey:2016mio}.

The Lagrangian in Eq.~\eqref{eq:GNLagrangian} is equipped with a discrete chiral symmetry
\begin{align}
    \psi \rightarrow \gamma_5\psi\,,\quad \qquad \bar{\psi} \rightarrow - \bar{\psi} \gamma_5\,. \label{eq:chiralsymm}
\end{align}
Further, the model can be formulated with global $O(N)$ or $SU(N/2)$ symmetry.
Notably, the $O(N)$ GN model has been investigated, recently, in conformal field theory \cite{Iliesiu:2015qra,Iliesiu:2017nrv} and in renormalization group calculations up to three-loop order~\cite{Gracey:1991vy}.
For the $SU(N/2)$ GN model, four-loop order renormalization group calculations are available~\cite{Gracey:2016mio} from which the $O(N)$ renormalization group functions can be recovered at that order.

The GN model possesses several instructive features which it shares with other
models from different areas of physics, including solid state physics and quantum chromodynamics.
Therefore, it serves as a laboratory to study ideas of more complicated systems in a lucid but comprehensive manner.
Importantly, in $2 < D < 4$ the GN model exhibits dynamical breaking of discrete chiral symmetry upon which the fermions acquire a finite mass in the true vacuum.
This corresponds to a continuous phase transition and at the corresponding renormalization group fixed point, the theory is believed to be an interacting conformal field theory.

\subsection{Gross-Neveu-Yukawa model}\label{sec:GNYmodel}

Performing a Hubbard-Stratonovic transformation or partial bosonization~(PB), the GN model can be reformulated introducing a single component real scalar field $\phi$ and a Yukawa interaction
\begin{align}
    \mathcal{L}_\text{PB} = \bar{\psi}_i \slashed{\partial} \psi_i  + \frac{1}{2}g\phi\bar{\psi}_i \psi_i - \frac{1}{2}\phi^2\,. \label{eq: GN HS}
\end{align}
The corresponding functional integral now has to be carried out over fermionic and bosonic fields.
Accordingly, we define the corresponding Yukawa theory also known as the Gross-Neveu-Yukawa (GNY) model, reading
\begin{align}
  \mathcal{L}_\text{GNY} = \bar{\psi}_i (\slashed{\partial} + \sqrt{y} \phi ) \psi_i + \frac{1}{2} \phi (m^2 - \partial^2) \phi + \lambda \phi^4\,. \label{eq:GNYLagrangian}
\end{align}
Here, the real scalar field $\phi$ has a canonical kinetic term and a quartic interaction  with coupling constant~$\lambda$.
Further, the fermions couple to the scalar with the Yukawa coupling~$\sqrt{y}$.
Notably, the GNY model is perturbatively renormalizable in $D=4-\epsilon$ dimensions.

The discrete chiral symmetry of the Gross-Neveu model, cf.~Eq.~\eqref{eq:chiralsymm}, is complemented by the corresponding transformation property of the scalar degree of freedom, i.e. the complete discrete chiral transformation for the GNY model reads
\begin{align}
    \psi \rightarrow \gamma_5\psi\,,\quad \qquad \bar{\psi} \rightarrow - \bar{\psi} \gamma_5\,, \qquad \phi \rightarrow -\phi\,.
\end{align}
This transformation leaves the GNY Lagrangian invariant and, in particular, it prevents the presence of a finite cubic scalar interaction term.
%

\subsection{Renormalization group procedure}\label{sec:RGmethod}
%
Both models, i.e. the GN model and the GNY model, were analyzed by the
perturbative renormalization group approach employing dimensional
regularization and the modified minimal subtraction scheme $\overline{\text{MS}}$
to four-loop order~\cite{PhysRevD.96.096010,Gracey:2016mio}.
In this section we will briefly review our RG analysis for Gross-Neveu-Yukawa
model \cite{PhysRevD.96.096010} in order to fix the notation.
For details of the renormalization group analysis of the Gross-Neveu model we refer to Ref.~\cite{Gracey:2016mio}.

The bare Lagrangian of the GNY model is defined based on Eq.~\eqref{eq:GNYLagrangian} upon replacing fields and couplings by their bare
counterparts $x \rightarrow x_0$ for $x\in\{\psi,\phi,y,\lambda,m\}$.
The renormalized Lagrangian then reads
\begin{align}
  \mathcal{L}_\text{GNY} &= Z_\psi\bar\psi\slashed{\partial}\psi-\frac{1}{2}Z_\phi(\partial_\mu\phi)^2+Z_{\phi^2}\frac{m^2}{2}\phi^2 \\
  &\qquad +Z_{\phi\bar\psi\psi} \sqrt{y} \mu^{\epsilon/2}\phi\bar\psi \psi+Z_{\phi^4}\lambda\mu^\epsilon(\phi\phi)^2\,, \nonumber
\end{align}
where $\mu$  defines the energy scale which parameterizes the RG flow of the couplings.
Here, the wave function renormalization constants $Z_\psi$ and $Z_\phi$ were defined, which relate the bare and the
renormalized Lagrangian upon the field rescaling $\psi_0 = \sqrt{Z_\psi} \psi$ and $\phi_0 = \sqrt{Z_\phi} \phi$.
The explicit dependence on the renormalization scale  $\mu$ in the lagrangian reflects that, after integration
over $D = 4-\epsilon$ dimensional spacetime, the couplings are shifted to render them dimensionless.
The renormalization constants for the mass term, the Yukawa coupling and the quartic coupling are introduced by
\begin{align}
 m^2 &= m_0^2 Z_\phi Z_{\phi^2}^{-1}, \\[5pt]
 y &= y_0 \mu^{-\epsilon} Z_\psi^2 Z_\phi Z_{\phi\bar{\psi}\psi}^{-2}, \\[5pt]
 \lambda &= \lambda_0 \mu^{-\epsilon} Z_\phi^2 Z_{\phi^4}^{-1}\,.
\end{align}
These relations provide the RG scale dependence of the renormalized quantities.
%

\subsection{Renormalization group functions}\label{sec:RGfunctions}

From the renormalization constants we construct the RG beta and gamma
functions, which denote the corresponding logarithmic derivatives of the
renormalization constants, i.e.
\begin{align}
\beta_x & = \frac{\mathrm{d}\, x}{\mathrm{d}\ln \mu}\,,\quad \mathrm{for}\quad x\in\{  y, \lambda \}\,,\\[5pt]
\gamma_x &= \frac{\mathrm{d}\ln Z_x}{\mathrm{d} \ln \mu}\,,\quad \mathrm{for}\quad x \in \{ \psi,\phi, \phi^2 \}\,.
\end{align}
We employ rescaled couplings $x/(8\pi^2) \rightarrow x$ for $x\in\{ y, \lambda \}$.
The renormalization group functions for the GNY model then have the form
\begin{align}
  \beta_x &=  -\epsilon\, x +  \sum\limits_{k=1}^{L} \beta_{x}^{(k\text{L})}\,, \quad \text{with} \quad x \in \{y, \lambda\}\,,\\
  \gamma_x &= \sum\limits_{k=1}^{L} \gamma_{x}^{(k\text{L})}\,, \quad \text{with} \quad x \in \{ \psi,\phi, \phi^2 \}\,.
\end{align}
At one-loop order, they explicitly read
\begin{align}
   \beta_{y}^{(\text{1L})} &= (3+N/2) y^2 \,, \label{eq:betay1L} \\
   \beta_{\lambda}^{(\text{1L})} &= 36 \lambda^2 + N y \lambda - (N/4)y^2  \,. \label{eq:betalambda1L}
\end{align}
and
\begin{align}
	\gamma_\psi^{(1\mathrm{L})}&=\frac{y}{2}\,,\quad \gamma_\phi^{(1\mathrm{L})}=2Ny\,,\quad \gamma_{\phi^2}^{(1\mathrm{L})}=-12\lambda\,.
\end{align}
The full expressions for the RG beta and gamma functions of the Gross-Neveu-Yukawa model up to four-loop order can be found in our Ref.~\cite{PhysRevD.96.096010} and in App.~\ref{app:betagamma}.

\subsection{Fixed points and critical exponents}

At a renormalization group fixed point the system becomes scale invariant which can be related to the occurrence of universal critical behavior near a continuous phase transition.
The beta functions allow the determination of the RG fixed points order by order in $\epsilon$.
For example, using the one-loop beta functions from Eqs.~\eqref{eq:betay1L} and~\eqref{eq:betalambda1L} one finds a physical and stable non-Gau\ss ian fixed point with the fixed point couplings
\begin{align}
   (y_*, \lambda_*) &= \left( \frac{\epsilon}{3+N/2}, \frac{(3-N/2+s)\epsilon}{72(3+N/2)}  \right)\,,\label{eq:1loopfp}
\end{align}
where $s = \sqrt{9+N(33+N/4)}$. Accordingly, the RG fixed point of the GN model in the coupling constant $g$ can be determined~\cite{Gracey:2016mio}.

As for the universal critical exponents, we are interested in the (inverse) correlation-length exponent $\nu^{-1}$ and the anomalous dimensions of bosons and fermions $\eta_\phi$ and $\eta_\psi$.
The latter ones are obtained by evaluating the renormalization group functions $\gamma_\phi$ and $\gamma_\psi$
at the fixed point, i.e.
\begin{align}
  \eta_\phi &= \gamma_\phi(y_*,\lambda_*)\,,
  \eta_\psi &= \gamma_\psi(y_*,\lambda_*)\,.
\end{align}
The inverse correlation length exponent $\nu^{-1}$ can be extracted from the relation
\begin{align}
 \nu^{-1} = 2 - \eta_\phi + \eta_{\phi^2}\,,
\end{align}
where, in agreement with the previous notation, we have defined $\eta_{\phi^2} = \gamma_{\phi^2}(y_*,\lambda_*)$.

The resulting epsilon expansion can be performed for a general number of spinor components $N$ to order $\mathcal{O}(\epsilon^4)$.
Schematically, the epsilon expansions for a critical exponent to order $L$ reads
\begin{align}\label{eq:expgen}
      f^{\mathrm{GN(Y)}}(\epsilon) = \sum\limits_{k=0}^{L} f_k^{\mathrm{GN(Y)}}\epsilon^k\,,
\end{align}
where $f$ represents a critical exponent, i.e. for the present discussion $f \in \{ \nu^{-1}, \eta_\phi, \eta_\psi \}$. As indicated, this also holds accordingly for the $2+\epsilon$ expansion of the GN~model. The full expressions for expansions of the critical exponents of the GNY~model in $D=4-\epsilon$ can also be calculated analytically, see the ancillary files of Ref.~\cite{PhysRevD.96.096010} and for the GN~model in $2+\epsilon$ dimensions in Ref.~\cite{Gracey:2016mio}.\medskip

\begin{figure*}[htbp]
    \includegraphics[scale=1.0]{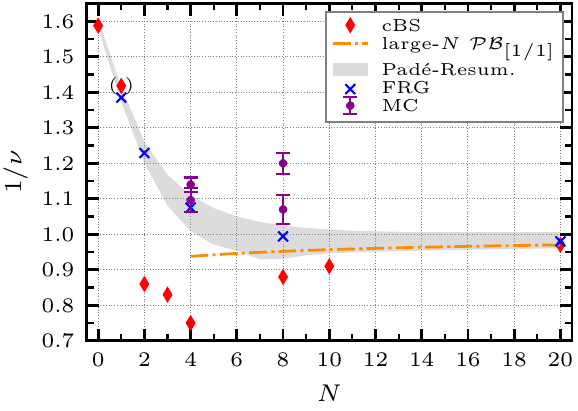}
    \includegraphics[scale=1.0]{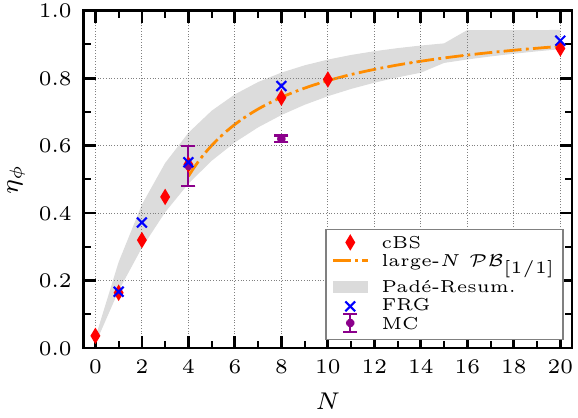}
    \includegraphics[scale=1.0]{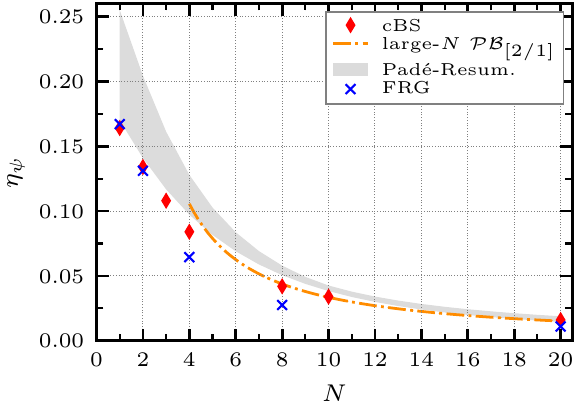}
    \caption{\textit{Chiral Ising universality} in $D=3$: Overview plots for the correlation-length exponent $\nu^{-1}$ (left panel), the boson anomalous
     dimension $\eta_\phi$ (medium panel) and the fermion anomalous dimension $\eta_\psi$ (right panel) for different numbers of spinor components $N\in[1,20]$.
     For comparison, values from Monte Carlo (MC) calculations, the functional renormalization group (FRG), and the conformal bootstrap (cBS) are also shown.
     For the large-$N$ results, we applied Pad\'e-Borel resummation.}
    \label{fig:criticalexponentsN}
\end{figure*}

Here, we explicitly display the series for $N=8$, which we study exhaustively in the following. The $4-\epsilon$ expansion of the GNY model gives the critical exponents
\begin{align}
	\frac{1}{\nu} =& 2-\frac{20\epsilon}{21}+\frac{325\epsilon^2}{44982}-\frac{(271572144\zeta_3+36133009)\epsilon ^3}{3821940612}\nonumber\\
	&-\frac{\left(2472257012904 \pi ^4+86141171013035\right)}{4175164363361040}\epsilon ^4\nonumber\\
	&+\frac{5 (70350676 \zeta_3+172549473 \zeta_5) }{20065188213}\epsilon^4+\mathcal{O}(\epsilon^5)\,,
	\\[5pt]
	\eta_\phi =& \frac{4 \epsilon }{7}+\frac{109 \epsilon ^2}{882}+\left(\frac{1170245}{26449416}-\frac{144 \zeta _3}{2401}\right) \epsilon ^3\nonumber\\
	&+\left(\frac{20491307339}{481564517112}-\frac{6 \pi ^4}{12005}\right) \epsilon ^4\nonumber\\
	&+\frac{4 (390883 \zeta_3+413100 \zeta_5) }{23143239}\epsilon ^4+\mathcal{O}(\epsilon^5)\,,\\[5pt]
	\eta_\psi =& \frac{\epsilon }{14}-\frac{71 \epsilon ^2}{10584}-\left(\frac{18 \zeta _3}{2401}+\frac{2432695}{158696496}\right) \epsilon ^3\nonumber\\
	&-\frac{\left(3610229616 \pi ^4+556332486445\right)}{57787742053440}\epsilon^4\nonumber
	\\
	&+\frac{(11109323 \zeta_3+4957200 \zeta_5) }{555437736}\epsilon ^4+\mathcal{O}(\epsilon^5)\,.
\end{align}
Here $\zeta_z$ is the Riemann zeta function. For other values of $N$, the critical exponents can be determined from the RG beta and gamma functions listed in App.~\ref{app:betagamma} or from Ref.~\cite{PhysRevD.96.096010}.
The $2+\epsilon$ expansion of the GN model can be found in Ref.~\cite{Gracey:2016mio} has the critical exponents
\begin{align}
	\frac{1}{\nu}&=\epsilon-\frac{1}{6}\epsilon^2-\frac{5}{72}\epsilon^3+\frac{81\zeta_3+35}{216}\epsilon^4+\mathcal{O}(\epsilon^5)\,,\\
	\eta_\phi&=2-\frac{4}{3}\epsilon-\frac{7}{36}\epsilon^2+\frac{7}{54}\epsilon^3+\frac{1092\zeta_3+91}{5184}\epsilon^4+\mathcal{O}(\epsilon^5)\,,\nonumber\\
	\eta_\psi&=\frac{7}{72}\epsilon^2-\frac{7}{432}\epsilon^3+\frac{7}{10368}\epsilon^4+\mathcal{O}(\epsilon^5)\,.
\end{align}
For general $N$, we have listed the critical exponents extracted from Ref.~\cite{Gracey:2016mio} in App.~\ref{app:JohnGN}.

\section{Pad\'e approximants}\label{sec:Pade}

\begin{figure*}[t!]
      \includegraphics[scale=1.0]{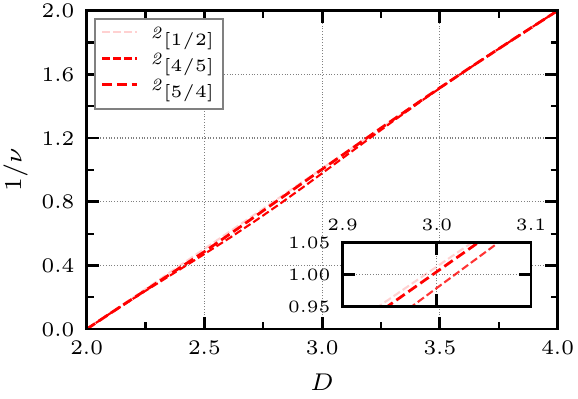}
      \includegraphics[scale=1.0]{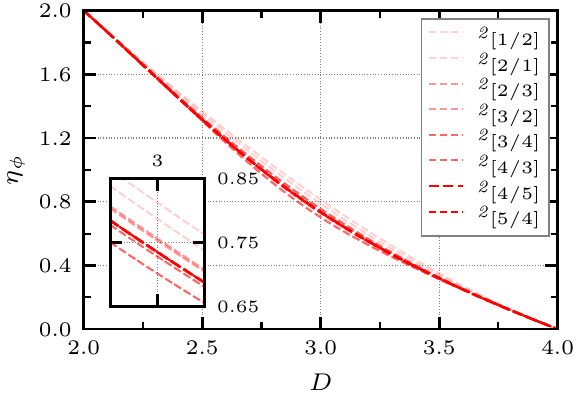}
      \includegraphics[scale=1.0]{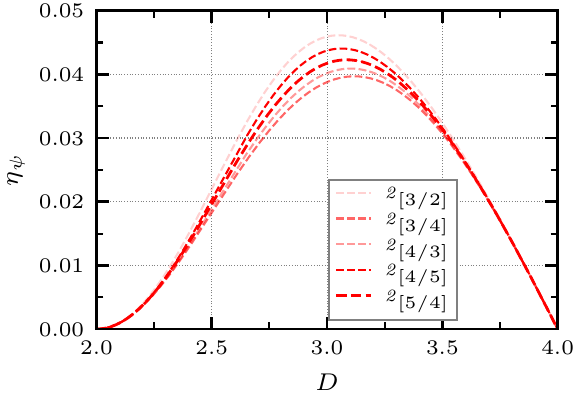}
      \caption{Order by order study of two-sided Pad\'e approximants of the three critical exponents $\nu^{-1}$ (left panel), $\eta_\phi$ (mid panel) and $\eta_\psi$ (right panel) between $D \in (2,4)$ at $N = 8$. We restrict ourselves to approximants, which include the same number of loop orders in the respective critical dimensions $D=2$ and $D=4$. Further, we choose approximants with $m \approx n$.}
      \label{fig: two-sided Pade}
\end{figure*}

We start our analysis of the Ising Gross-Neveu critical exponents by studying  Pad\'e approximants of the exponents' series expansions $f(\epsilon)=\sum_{k=0}^L f_k \epsilon^k$ up to order $L$. The corresponding Pad\'e approximants are rational functions
\begin{align}\label{eq:padegen}
      [m/n] = \frac{a_0 + a_1 \epsilon + \dots + a_m \epsilon^m}{1+b_1 \epsilon + \dots + b_n \epsilon^n} \,.
\end{align}
with $L=m+n$ where the expansion coefficients $a_0,..., a_m, b_1,..., b_n$ are uniquely given by the condition that the series expansion of $[m/n]$ match the original series
\begin{align}
  [m/n] - \sum\limits_{k=0}^L f_k \epsilon^k = \mathcal{O}(\epsilon^{L+1}) \,.
\end{align}
At a given order $L$, there are $L+1$ different Pad\'e approximants and we
note that
 it is not \emph{a priori} clear which of them
will give the most faithful estimate.
Importantly, Pad\'e approximants can be used for finding approximations to
functions outside the radius of convergence $R$ of their corresponding power
series. In particular,  the case $R=0$ is relevant in the context of the perturbative RG~\cite{Kleinert, Baker1975}.
Further, no assumption about the large-order behavior of the series coefficients is made.

We employ this method for the critical exponents from the $(4-\epsilon)$ expansion
series of the GNY model at four-loop order by evaluating all $4+1$ Pad\'e approximants at $\epsilon=1$, i.e. for $D=2+1$ dimensions.
This provides a range of estimates for the critical exponents, which we interpret as a first rough window of confidence for their values.
We show the results of this analysis as a function of $N$ in Fig.~\ref{fig:criticalexponentsN} as the gray-shaded area.
There, we have only taken into account Pad\'e approximants which show no poles in the range $D\in(2,4)$.
Since this criterion is not fulfilled by all approximants for a given $N$, a
sudden enlargement of the window may occur upon changing $N$.
 Such an  example can be seen for the  correlation length exponent $\nu^{-1}$
around $N\approx~7$. In this case, the pole of one of the approximants is pushed out of the interval $D\in(2,4)$.
In the following section, when we consider two-sided Pad\'e approximants, we will also show sequences of Pad\'e approximants to study more carefully the convergence properties of the approximations.

For comparison, we also present the results of other methods for different $N$ in Fig.~\ref{fig:criticalexponentsN}.
The functional renormalization group (FRG)~\cite{Vacca2015,Gies2017,PhysRevB.94.245102} provides compatible estimates for $\nu^{-1}$ and $\eta_\phi$ over the whole range of $N$.
On the other hand, the FRG values for the fermionic anomalous dimension $\eta_\psi$ are systematically below the range we find with Pad\'e approximants.
We remark that the FRG calculations in Refs.~\cite{Vacca2015,Gies2017,PhysRevB.94.245102} are based on the derivative expansion scheme and more intricate momentum dependencies might become important for the evaluation of anomalous dimensions~\cite{PhysRevE.80.030103,PhysRevE.85.026707}.
Comparison to the conformal bootstrap results from Ref.~\cite{Iliesiu:2017nrv} also shows very good compatibility concerning the boson anomalous dimension.
Generally, the results for the fermion anomalous dimension lie between the FRG estimates and the window given by the Pad\'e approximants.
We show in the next two sections, how this discrepancy can be resolved.
The conformal bootstrap results for the correlation length exponents deviate strongly from the other approaches, in particular, in the range $2 \leq N \leq 8$.
For very small $N$, including the limit of the Ising universality class and the emergent SUSY limit $(N=1)$, the estimates agree well with the renormalization group approaches.
We note that within the conformal bootstrap method, the boson and fermion anomalous dimensions are obtained from universal bounds. The correlation length exponent estimate, however, is based on the extremal functional approach where one assumes that the theory exactly lives at the characteristic kink and subsequent extrapolation of the spectrum~\cite{Iliesiu:2017nrv}.
Therefore, it would be very interesting to obtain conformal bootstrap estimates of the correlations length exponent directly from universal bounds.
From Monte Carlo simulations estimates for the Ising Gross-Neveu universality class are available for the cases $N=4$ and $N=8$.
In particular, the case $N=4$ was intensely studied by Monte Carlo methods~\cite{1367-2630-16-10-103008,1367-2630-17-8-085003,PhysRevB.93.155157,PhysRevD.96.114502,Schmidt:2017} and the resulting estimates still varied with system size. In Fig.~\ref{fig:criticalexponentsN}, we show the latest estimates for the correlation length exponent from Refs.~\cite{PhysRevD.96.114502,Schmidt:2017} for comparison, which agree quite well with the RG estimates, but not with the conformal bootstrap.
For the boson anomalous dimension, the latest estimate from Ref.~\cite{PhysRevD.96.114502} is also in good agreement with the other approaches.
For the fermionic anomalous dimension, only a value at $N=8$ is available~\cite{Chandrasekharan:2013aya}, which with $\eta_\psi \approx 0.38(1)$ is much larger than the estimates from the complementary methods and therefore does not appear in the range presented in the Fig.~\ref{fig:criticalexponentsN}. Also the $N=8$ estimates from Refs.~\cite{Chandrasekharan:2013aya,PhysRevB.97.081110} for the correlation length exponent and the boson anomalous dimension are not in agreement with RG or the conformal bootstrap. In the left panel of Fig.~\ref{fig:criticalexponentsN}, we also show the recent results from Ref.~\cite{Schmidt:2017} for the $N=8$ correlation length exponent, which is much closer to our result than the one from Ref.~\cite{Chandrasekharan:2013aya}.
Finally, as another perturbative method we also show the values from a large-$N$ expansion \cite{Karkkainen:1993ef,Janssen2054} for $N\ge 4$.
Note that we have resummed these series following reference \cite{Karkkainen:1993ef} using the Pad\'e-Borel method.
The resulting curves are for most $N$ deep in the shaded area of the Pad\'e approximants.

\begin{figure*}[htbp]
      \includegraphics[scale=1.0]{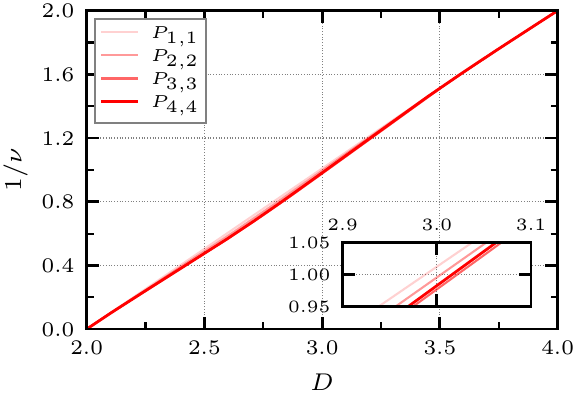}
      \includegraphics[scale=1.0]{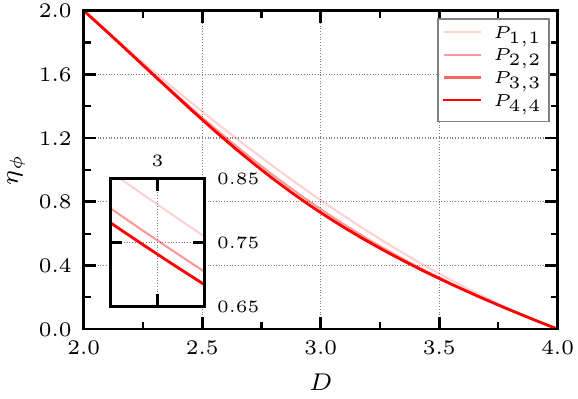}
      \includegraphics[scale=1.0]{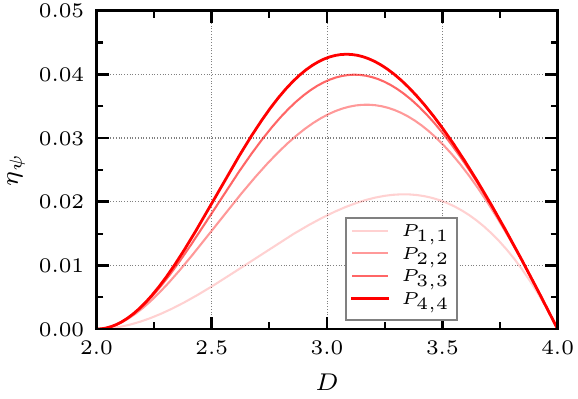}
      \caption{Order by order study of polynomial interpolation of the three critical exponents $\nu^{-1}$ (left panel), $\eta_\phi$ (mid panel) and $\eta_\psi$ (right panel) between $D \in (2,4)$ at $N = 8$. We restrict ourselves to interpolations, which include the same number of loop orders in the respective critical dimensions $D=2$ and $D=4$}
      \label{fig:interpolationD}
\end{figure*}

\subsection{Two-sided Pad\'e approximants}

In Sec.~\ref{sec:models}, we have introduced the Gross-Neveu model, which is expanded in $2+\epsilon$ dimensions and the Gross-Neveu-Yukuwa model expanded $D=4-\epsilon$ dimensions.
These models are closely related through their symmetries and symmetry-breaking patterns and universality therefore suggests that they describe the same critical point~\cite{ZINNJUSTIN1991105}.
Moreover, the $2+\epsilon$ and the $4-\epsilon$ expansions can be compared to the known results from $1/N$ expansions~\cite{Gracey:1990wi,Gracey:1992cp,Vasiliev:1992wr,Vasiliev:1993pi,Gracey:1993kb,Gracey:1993kb,Gracey:1993kc,Derkachov:1993uw,Gracey:2017fzu,Manashov:2017rrx}.
Indeed, we have confirmed that the $4-\epsilon$ expansion is fully compatible with the large-$N$ expansion of the Gross-Neveu model order by order, which represents a highly nontrivial check, see our Refs.~\cite{Mihaila:2017ble,PhysRevD.96.096010}.

In the following, we combine the two expansion schemes, which are defined  near their respective critical dimensions, i.e. $D=2$ and $D=4$.
To that end, we again employ a generic Pad\'e-approximant,
cf.~Eq.~\eqref{eq:padegen}, and this time, we fix its coefficients such that
its power series expansion near $D=2$ and $D=4$ agrees with both perturbative
series  in $2+\epsilon$ and the $4-\epsilon$ dimensions, respectively. Explicitly, we make the ansatz
\begin{align}
      \textit{2}_{[m/n]}(D) = \frac{a_0 + a_1 D + \dots + a_m D^m}{1+ b_1 D + \dots + b_n D^n}\,,
\end{align}
and demand for the coefficients $\{a_i\}$ and $\{b_j \}$ to fit to the epsilon expansions. This leads to the relations
\begin{align}
      \textit{2}_{[m/n]}^{(k)}(2) &= k! f_k^\text{GN}\,,\\
       \textit{2}_{[m/n]}^{(k)}(4) &= (-1)^k k! f_k^\text{GNY}\,,
\end{align}
with $f_k^{\text{GN(Y)}}$ being the expansion coefficient of a critical exponent at order $k$, cf.~Eq.~\eqref{eq:expgen}.
In this way, the two-sided Pad\'e approximant $\textit{2}_{[m/n]}(D)$  provides an interpolating function for a critical exponent $f(D)$ in $2<D<4$.

We show the two-sided Pad\'e approximants evaluated for the inverse correlation length exponent and the boson and fermion anomalous dimensions for the physically interesting case $N=8$ in Fig.~\ref{fig: two-sided Pade}.
Moreover, we also show a sequence of two-sided Pad\'e approximants corresponding to increasing order of the perturbative expansions which clearly shows signs of convergence towards higher orders.
The two-sided Pad\'e approximants can also have a pole in the interval $D\in(2,4)$ depending on the choice of $f$ and $N$.
In Fig.~\ref{fig: two-sided Pade}, we therefore show only the two-sided Pad\'e approximants, which do not have a pole in $D\in(2,4)$ for the example $N=8$.
We further restrict our analysis to approximants which include the same number of loop orders at both ends, i.e. at $D=2$ and at $D=4$.
We observe a very good stability of the estimates from the two-sided Pad\'e approximants for the three critical exponents upon increasing the orders of the  expansions.
In particular, this also holds for the fermion anomalous dimension which
vanishes in both limits, i.e. at $D=2$ and $D=4$ and is finite only in between.
Comparisons to other methods will be presented in the next section after we have also analysed an alternative interpolation method.
Finally, we remark that the series expansions for the critical exponents of the Gross-Neveu model have a pole at $N=2$ and therefore an extraction of estimates from the $2+\epsilon$ expansion close to or below $N=2$ becomes problematic.
Therefore, we find that two-sided Pad\'e approximants and other interpolation schemes cannot be faithfully applied for small $N$.
In fact, the effects of the pole at $N=2$ can already be observed at $N=4$.

\section{Polynomial interpolation}\label{sec:polyinterpolation}

\begin{figure*}[t!]
      \includegraphics[scale=1.0]{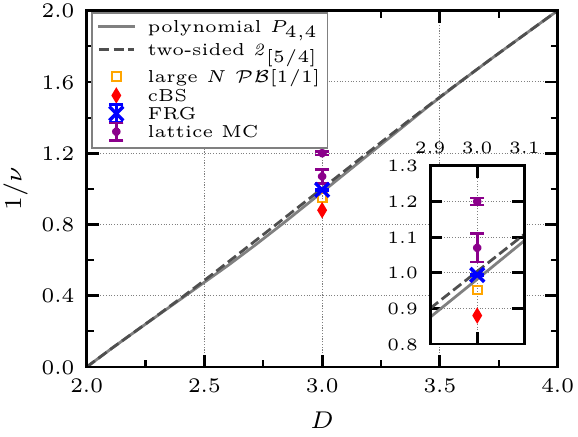}
      \includegraphics[scale=1.0]{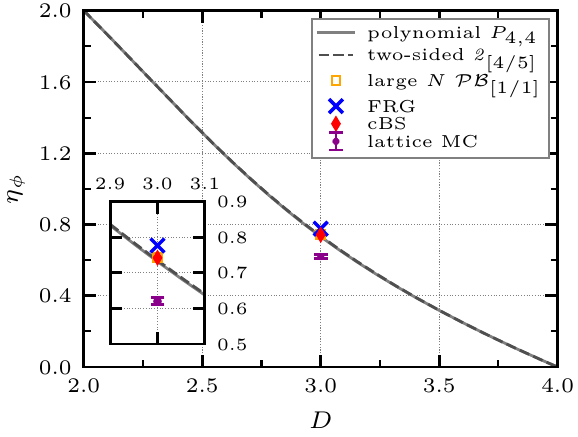}
      \includegraphics[scale=1.0]{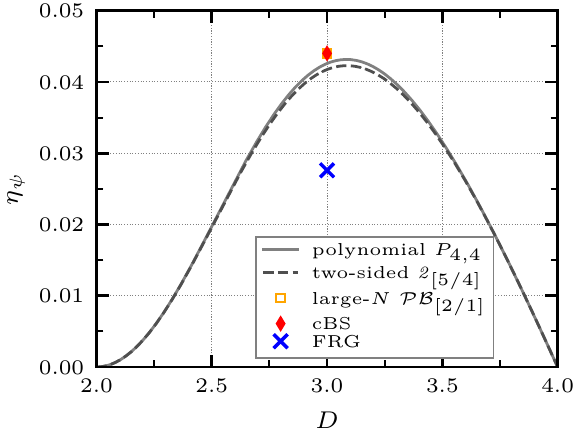}
      \caption{Interpolation of the three critical exponents $\nu^{-1}$ (inverse correlation-length exponent, left panel), $\eta_\phi$ (boson anomalous dimension, mid panel) and $\eta_\psi$ (fermion anomalous dimension, right panel)
      between $D \in (2,4)$ at $N = 8$. The shown interpolations polynomial (red line) and two-sided Pad\'e (dark-red dashed line) are fixed by the two epsilon-expansions at $D=$2 and $D=4$ in the first four derivatives. As a result,
      the asymptotic behavior is suppressed even far from these expansion points at the physical dimension $D=3$ and plausible values can be read off. It should also be noted that both complementary approaches
      for the interpolation are very close to each other and comparable to conformal bootstrap (cBS) \cite{Iliesiu:2017nrv}, lattice Monte Carlo \cite{PhysRevD.88.021701} and functional renormalization group (FRG) \cite{PhysRevB.94.245102} calculations.
      Also note that no point for the fermionic anomalous dimension has been plotted for MC, since the point with $\eta_\psi \approx 0.38(1)$ is far above the value range of the comparable methods.}
      \label{fig:interpolationD2}
\end{figure*}

A suitable interpolation between the two critical dimensions can also  be constructed by using a purely polynomial ansatz.
To that end, we use both epsilon expansions for the critical exponents, simultaneously, and, again, set up an interpolating function for the critical exponents in the interval $D\in(2,4)$.
More specifically, for an exponent $f(D)$, we choose a polynomial interpolation with polynomial $P_{i, j} (D)$ of $(i+j)$-th degree, where $i$ ($j$) denotes the highest order of the epsilon expansion in $D=2+\epsilon$ ($D=4-\epsilon$) dimensions.
We fix the polynomial coefficients with the expansion near the lower critical dimension and determine the first $i+1$ terms from the condition
\begin{align}
      P_{i,j}(D) = \sum\limits_{k=0}^{i} f_k^\text{GN} (D-2)^k + \sum\limits_{k=i+1}^{i+j+1}a_k (D-2)^k\,.
\end{align}
The remaining $(j+1)$ higher-order terms with coefficients $a_{i+1},\dots,a_{i+j+1}$ are then determined from the requirement that the $j$ lowest derivatives of
$P_{i, j} (D)$ at $D=4$ correspond to the $\epsilon$-expansion of the GNY model by
\begin{align}
      P_{i,j}^{(n)}(4) = (-1)^n n! f_n^\text{GNY}
\end{align}
where $P_{i, j}^{ (n)} (4)$ denotes the $n$-th derivative at $D=4$.
The resulting polynomials are then by construction $i$-loop ($j$-loop) exact near the lower (upper) critical dimension and provide a systematic estimate for the physically relevant case of three dimensions.

We show the results of this interpolation procedure, i.e. $P_{i,i}(D)$ for $i\in \{1,2,3,4\}$ for the
inverse correlation length exponent and the anomalous dimensions for the
case $N=8$ in Fig.~\ref{fig:interpolationD}.
Again, for the inverse correlation length exponent and the boson anomalous dimension, we observe a very good stability of the estimates from the interpolation procedure upon increasing the orders of the two perturbative expansions.
In contrast to the two-sided Pad\'e approximants, the estimates for the fermion anomalous dimension are less stable but still show signs of convergence when comparing polynomial interpolations from one order to the next for higher and higher orders.

In Fig.~\ref{fig:interpolationD2}, we summarize our best results for the $N=8$ estimates from the two-sided Pad\'e approximants as well as from the polynomial interpolation exhibiting the excellent agreement between both interpolations in the whole range $D\in (2,4)$.
For comparison, we also show the estimates for the critical exponents at $D=3$ from other methods, namely from the functional RG~\cite{PhysRevB.94.245102} and from the conformal bootstrap~\cite{Iliesiu:2017nrv}.
In particular, the conformal bootstrap estimates for the boson and fermion anomalous dimensions, which have been determined from universal bounds, almost perfectly match with our results.
There is still a sizable difference in the estimates of the inverse correlation length exponent, which needs to be resolved in future studies.
The available quantum Monte Carlo results for $N=8$~\cite{Chandrasekharan:2013aya,Schmidt:2017} show deviations from our RG results as well as from the conformal bootstrap estimates for both, the anomalous dimensions and the correlation length exponent. It is encouraging, though, that the more recent QMC results from Ref.~\cite{Schmidt:2017} seem to agree better than the earlier estimates from Ref~\cite{Chandrasekharan:2013aya}. Unfortunately, in Ref.~\cite{Schmidt:2017} only the correlation length exponent is given and the situation for the anomalous dimensions remains to be clarified.
We remark, that in the more exhaustively studied case of $N=4$, it has been found that the QMC results can still be subject to some changes upon increasing the lattice size~\cite{PhysRevD.96.114502}.

We compile our results from the two different interpolation techniques and the following resummations for $D=3$ as a function of $N$ in Fig.~\ref{fig:criticalexponentsNallmethods} in the appendix.
Generally, the two interpolation procedures provide highly compatible results for  $N\gtrsim 6$ and start to deviate from each other and the other methods for $N\lesssim 6$. This is expected since the interpolation makes use of the series expansion in $2+\epsilon$ which exhibits poles in the critical exponents for~$N=2$, see Fig.~\ref{fig:criticalexponentsNallmethods}.
We conclude that for $N\lesssim 6$ we cannot faithfully employ the interpolation techniques rooting in a $2+\epsilon$ expansion.
In the following section, we therefore explore Borel resummation for the $4-\epsilon$ expansion to obtain improved estimates for the Gross-Neveu universality classes at small $N$, in particular for smaller $N$ where interpolation between $2+\epsilon$ and $4-\epsilon$ is difficult.

\section{Borel resummation} \label{sec:Borelresummation}

A very accurate determination of critical exponents in $D=3$ from the $(4-\epsilon)$ expansion was achieved for $O(n)$~symmetric $\phi^4$ theories by using Borel resummation with conformal mapping~\cite{Kleinert,ZinnJustin:2002ru}, see e.g., Ref.~\cite{Kompaniets2017} for a recent study at six-loop order.
For this resummation technique, the large-order behavior of an asymptotic series is considered, which has been computed for scalar models in the minimal subtraction scheme~\cite{0305-4470-17-9-021,0305-4470-11-11-013}.
Unfortunately, for the Yukawa models considered, here, the precise large-order behavior is not known.
However, even with the knowledge of the large-order behavior as in the O($N$) symmetric scalar models, resummation is a delicate issue.
There, for example, the series written in terms of the coupling constant in fixed dimensions $D=2,3$ is known to be Borel summable~\cite{Eckmann1975,Magnen1977}, but the situation for the epsilon expansion remains unsettled.

Borel summability is therefore often taken as an assumption in the analysis of $O(N)$ symmetric scalar models~\cite{Vladimirov1979} and we will also do this, here.
In the following, we also make the additional assumption that the asymptotic behavior of the GNY model is qualitatively the same as the one from the scalar models, i.e. the epsilon expansion for the critical exponents follows a formal power series with factorially increasing coefficients, i.e.
\begin{align}
      f_k \sim (-a)^k \mathrm{\Gamma}(k+b+1) \approx  (-a)^k k! k^b \label{eq:largeordercoeff}
\end{align}
for large $k$.
The Borel transform of such an asymptotic series $f$ with expansion coefficients $f_k$ is calculated as
\begin{align}
      \mathcal{B}^b_f(\epsilon) := \sum\limits_{k=0}^\infty \frac{f_k}{\mathrm{\Gamma}(k+b+1)}\epsilon^k = \sum\limits_{k=0}^\infty B_k^b \epsilon^k\,.
      \label{eq:Boreltransform}
\end{align}
Consequently, the coefficients behave like $B_k^b \sim (-a)^k$ for large orders $k$  and therefore follow a geometric series which can be understood as a rational function with a pole at $\epsilon=-1/a$, i.e.
\begin{align}
      \mathcal{B}^b_f(\epsilon) \underset{\text{$k$ large}}{\sim} \sum\limits_{k} (-a)^k \epsilon^k = \frac{1}{1+a \epsilon}\,.
\end{align}
While the original series has a vanishing radius of convergence, we note that the Borel transform is analytic in a circle with $|\epsilon|<1/a$.
We now use the assumption that the considered series are Borel summable, i.e. that we can analytically continue the Borel transform to the positive real axis and that the Borel sum
\begin{align}
      \tilde{f}(\epsilon) := \int\limits_{0}^\infty \mathrm{d}t \, t^b \mathrm{e}^{-t} \mathcal{B}^{b}_f(\epsilon t)\,,\label{eq:Borelsum}
\end{align}
is convergent and gives the correct value of $f$ for $\epsilon=1$.

A perturbative expansion of the integral in Eq.~\eqref{eq:Borelsum} with respect to $\epsilon$ restores the original asymptotic series.
In particular, if only finitely many terms of the asymptotic series are known, the corresponding finite Borel transform in the Borel sum merely reproduces the initial series.
This can be circumvented by replacing the Borel transform in Eq.~\eqref{eq:Borelsum} by a nonpolynomial function which has the same power series coefficients as the Borel transform for all known terms and ideally incorporates the large-order behavior of the series.
Therefore, in the next step, we analytically continue the Borel transform with a conformal transformation to the real axis, using
\begin{align}
  w(\epsilon) = \frac{\sqrt{1+a \epsilon} -1}{\sqrt{1+ a \epsilon} +1} \quad \Rightarrow \quad \epsilon = \frac{4}{a}\frac{w}{(1-w)^2}\,.
  \label{eq: conf mapping}
\end{align}
This transformation preserves the origin and maps all points of the relevant positive real $\epsilon$~axis in the unit circle, i.e. $|w(\epsilon)|<1$ for $\epsilon\in[0,\infty)$.
The cut of the negative real axis $(-\infty, -1/a)$ of the Borel transform is mapped to the unit circle in the variable $w$.
The Borel transform in the new variable $w$ can be found by expanding
$\mathcal{B}^b_f\left(\epsilon(w)\right)$ in $w$, which renders the series
convergent  in the full
integration domain of the Borel sum.

We further introduce an adjustment parameter $\lambda$ in the Borel transform truncated at order $L$, cf. Refs.~\cite{Kompaniets2017,Vladimirov1979},
\begin{align}
      \mathcal{B}^b_f(\epsilon) \approx \mathcal{B}^{a,b,\lambda}_f(\epsilon) := \left( \frac{a \epsilon(w)}{w} \right)^\lambda \sum\limits_{k=0}^L B_{k}^{b,\lambda} w^k\,.
      \label{eq:Borelsumlambda}
\end{align}
The coefficients $ B_{k}^{b,\lambda}$ are found by expanding the expression $(w/a\epsilon(w))^\lambda \mathcal{B}^b_f(\epsilon(w))$ in a power series of $w$.
While the Borel transform in Eq.~\eqref{eq:Boreltransform} for finite $L$ will diverge, 
this behavior is lost when introducing the conformal mapping variable as $w$ tends to one for large $\epsilon$.
The parametrization in Eq.~\eqref{eq:Borelsumlambda} with $\lambda \neq 0$ can  be used to restore the actual large $\epsilon$ behavior $\sim\epsilon^\lambda$ of the  Borel transform by adjusting $\lambda$.
Since the asymptotic behavior of $f(\epsilon)$ is unknown, we use $\lambda$ to
improve the sensitivity properties  of our resummation
algorithm~\cite{Vladimirov1979}, see below.

Further improvement is introduced by a homographic transformation with shifting variable  $q$
\begin{align}
      \epsilon = h_q(\tilde{\epsilon}) := \frac{\tilde{\epsilon}}{1+q \tilde{\epsilon}} \quad \Rightarrow \quad \tilde{\epsilon} = h^{-1}_q(\epsilon) = \frac{\epsilon}{1-q \epsilon}\,.
\end{align}
The original series is expanded in $\tilde{\epsilon}$ and in the last step before the Borel sum transformed back by $h^{-1}_q$.
Thus, the resummed series is determined by
\begin{align}
      \tilde{f}(\epsilon) \approx f_L^{a,b,\lambda,q}(\epsilon) := \int\limits_{0}^\infty \mathrm{d}t \, t^b \mathrm{e}^{-t} \mathcal{B}^{a,b,\lambda}_{f\circ h_q}(h^{-1}_q(\epsilon) t)\,.
      \label{eq: resummed series}
\end{align}
The conformal mapping in the Borel transform produces poles in the Borel sum Eq.~\eqref{eq:Borelsum}.
Using the homographic transformation these poles are shifted away from the physical dimension at $\epsilon=1$ into a region of the integral where their diverging behavior is suppressed by $\mathrm{exp}(-t)$.
This improves the stability of the numerical analysis and is essential to find an ``optimized'' parameter set.

\begin{figure}[t]
      \includegraphics[scale=0.702]{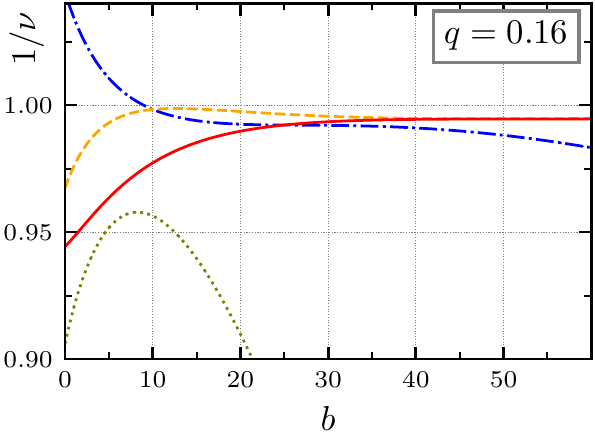}
      \includegraphics[scale=0.702]{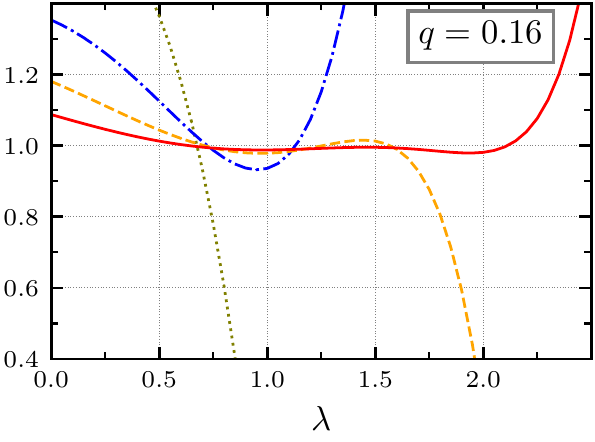} \\
      \includegraphics[scale=0.702]{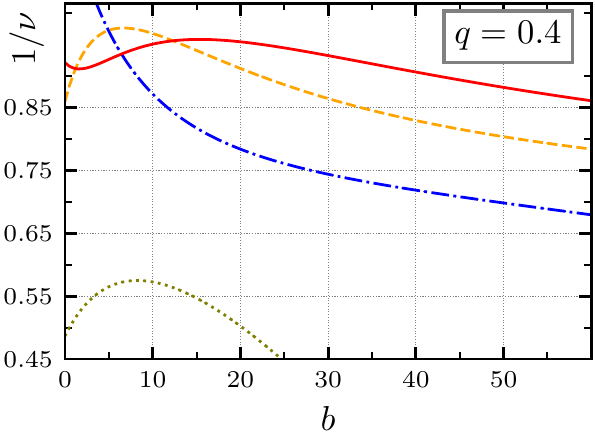} 
      \includegraphics[scale=0.702]{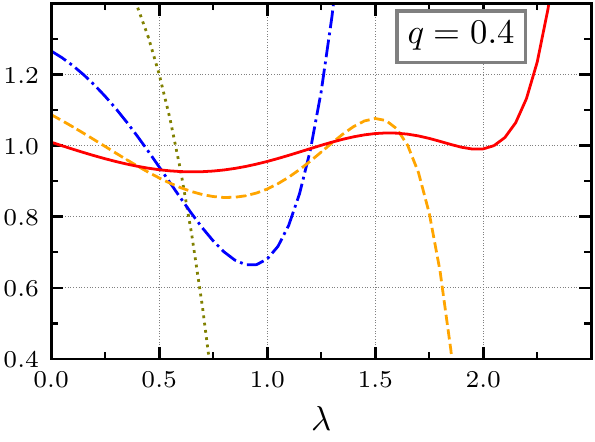} \\[-0.13cm]
      \hspace{-0.025cm}
      \includegraphics[scale=0.702]{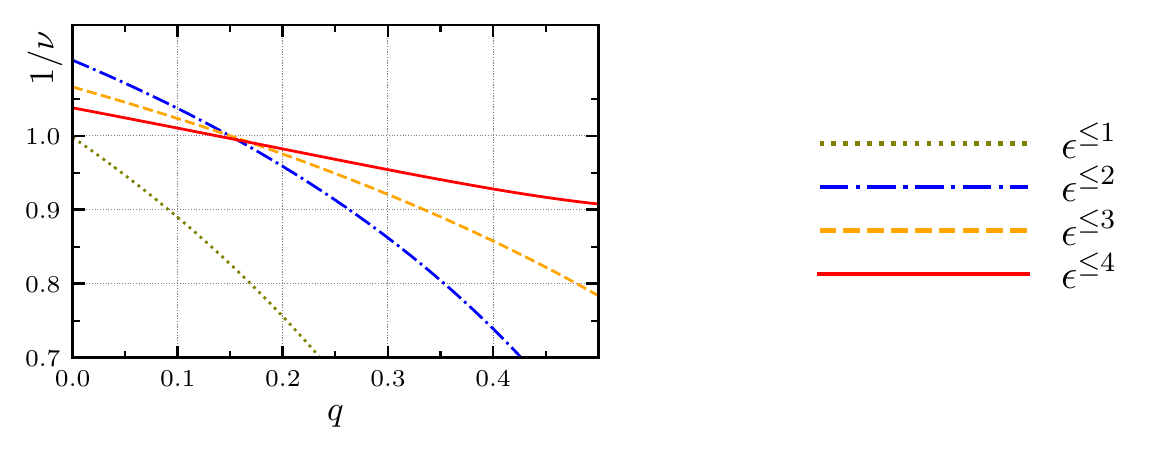} \hfill \ \vspace{-0.3cm}
      \caption{\textit{Chiral Ising universality} for $D=3, N=8$: Sensitivity of the Borel resummed inverse correlation-length exponent $\nu^{-1}$ to a variation of one of the three
        large-order parameters $b$, $\lambda$ and $q$. In the first row we show the change to variation of $b$ and $\lambda$ at the optimal $q=0.16$, i.e.
         at the global minimum of the error estimate $\overline{E} \equiv E_4(31.5,0.74,0.16) \approx 0.009$. In this case we find plateaus with increasing spread in the
         loop order. For comparison, we also show the same plots at $q=0.4$ away from the optimum where these plateaus are lost.
         In the last plot we present the variation with respect to $q$ which has no plateaus at all is only fixed by the intersection point.
          For a list of the minima of various $N$ configurations in the
        Borel resummation see Tab.~\ref{tab: minima} in the appendix.}
      \label{fig: Variation parameters Nuinv N8}
\end{figure}

To summarize, we introduced a resummation scheme for the asymptotic series $f(\epsilon)$ which incorporates four parameters $a$, $b$, $\lambda$ and $q$ in the resummed series $\tilde{f}(\epsilon)$ in Eq.~\eqref{eq: resummed series}.
The resummed series would be independent of the choice of these parameters, if all orders of the expansion were known, i.e. for $L \rightarrow \infty$.
At finite $L$ this suggests that, for an optimized choice of resummation parameters, their variation should only induce a mild variation of the resummed $\tilde{f}(\epsilon)$.
Consider, for example, the inverse correlation length exponent $1/\nu$, resummed according to Eq.~\eqref{eq: resummed series}.
Individually changing the parameters $b,\lambda$ and $q$ leads to a variation of the resummed $1/\nu$, which we show in Fig.~\ref{fig: Variation parameters Nuinv N8} for different orders in the epsilon expansion.
We observe that for increasing order in epsilon, the dependence on the resummation parameters generally decreases as expected.
Further, there are extended plateaus where the function $1/\nu$ of the resummation parameters becomes rather flat, i.e. the
resummed series becomes insensitive to a further variation in the corresponding parameter.

In the plots we omitted the dependence on the parameter
$a$, cf. Eq.~\eqref{eq:largeordercoeff}, which we fix to the value $a = (3+N/2)^{-1}$.
This is motivated by the observation that the factor $ (3+N/2)^{-1}$ structurally appears in the RG beta functions, fixed point values and critical exponents at all available loop orders, see, e.g., Eq.~\eqref{eq:1loopfp}.
This is reminiscent of the factor $3/(n+8)$ appearing in the study of the scalar $O(n)$ models, which determines $a$ in the corresponding asymptotic behavior~\cite{PhysRevD.15.1544}.
In our case, however, we do not have further justification to use a specific value for~$a$. We have explicitly checked the stability of our results with respect to variations of $a$ and we find that choosing different $a \in (0,1]$ only very mildly affects our results. For simplicity, we therefore proceed with the fixed choice $a = (3+N/2)^{-1}$.

\subsection{Resummation algorithm and error estimate}

We now describe the employed resummation algorithm closely following the strategy presented in Ref.~\cite{Kompaniets2017} and then apply it to the
epsilon expansion of the GNY model.
In order to find an optimized set of resummation parameters, we need a measure for the sensitivity of the summation to a parameter change.
To that end, we study the variation of a function $F$ upon changing  one of its parameters $x$ in a range $x\in[x_0, x_0+\Delta]$,
\begin{align}
      \mathcal{S}_x(F(x_0)) := \min\limits_{x\in[x_0,x_0+\Delta]}\left( \max\limits_{x'\in[x_0,x_0+\Delta]} \left| F(x) - F(x') \right| \right)\,.
      \label{eq: Sensitity function}
\end{align}
A function $F(x)$ which only weakly varies within the plateau of length $\Delta$ leads to a small value for $\mathcal{S}_x$.

Considering the example of the inverse correlation length $1/\nu$ in Fig.~\ref{fig: Variation parameters Nuinv N8} again, now, we are able to quantify the sensitity
by reading off the length $\Delta$ of the extended plateaus. At the highest available order $\mathcal{O}(\epsilon^4)$, these have the  spreads
\begin{align}
       \Delta_b \approx 40,  \quad        \Delta_\lambda \approx 1,     \quad     \Delta_q \approx 0.02\,, \label{eq: plateau length}
\end{align}
which we will use in the following analysis for computing the sensitivities in Eq.~\eqref{eq: Sensitity function}.
We note, that finding these plateaus without prior knowledge of the resummation parameters can be quite tedious since the parameter space is large.
Here, we have performed extensive scans and in Fig.~\ref{fig: Variation
  parameters Nuinv N8}, we already show a range of parameters for $b,\lambda$
and $q$ which is centered around the optimized set to be determined as
indicated in the caption. We want to stress at this point that the
  parameter $q$ is chosen in such a way that the dependence on the parameters
  $b$ and $\lambda$ exhibits extended flat plateaus. The  dependence on $q$ itself
  is rather steep over the whole definition domain and   we  use it just as an
  optimization tool for the numerical
  analysis~\cite{0305-4470-31-40-006}. This is also the reason why the
  variation domain for $q$ is very narrow. Note that the variation with respect to $q$ shows no plateaus and the optimal $q$ is fixed
  by the intersection point of the curves corresponding to the different loop orders, see lower left panel of Fig.~\ref{fig: Variation
  parameters Nuinv N8}.
\begin{figure}[t]
      \centering
      \includegraphics[scale=0.97]{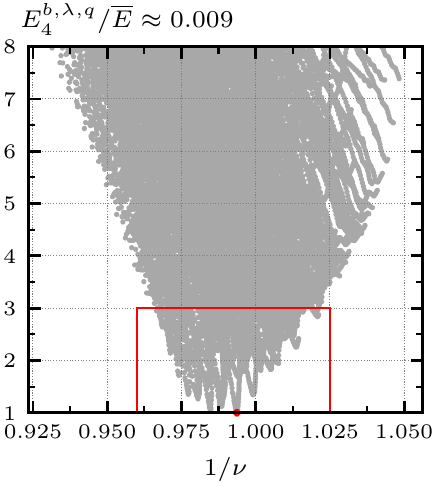}
      \includegraphics[scale=0.97]{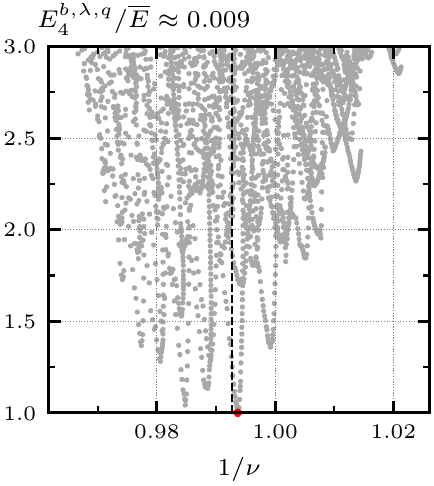}
      \caption{\textit{Chiral Ising universality} in $D=3$: Error estimates
        for the inverse correlation-length exponent $\nu^{-1}$ at  $N=8$. The
        error axis is normalized
      to the smallest error in the parameters $\overline{E} \equiv
      E_4(31.5,0.74,0.16) \approx 0.009$. The global minimum of this parameter
      space (left panel) is located in an area around the value $\nu^{-1}
      \approx 0.994$, which we have zoomed in on the right panel.
     The weighted mean of this subspace in parameter space has $\nu^{-1}\approx 0.993$ (dashed line).}
      \label{fig: scan nuinv N8}
\end{figure}

We identify the optimal set of parameters following the method introduce in
Ref.~\cite{Kompaniets2017}, i.e.  making use of both  the ``principle of
minimal sensitity'' and ``principle of fastest convergence''. As already
extensively discussed in the literature, the errors might be underestimated if
the unknown higher order corrections are much larger than the last computed
ones. Such examples are known and only an explicit calculation can solve this question.

The different variations are collected in an error estimate~\cite{Kompaniets2017},
\begin{align}
      E_{f,L}^{b,\lambda,q} := &\max\left\{ |f_{L}^{b,\lambda,q}-f_{L-1}^{b,\lambda,q}|, |f_{L}^{b,\lambda,q}-f_{L-2}^{b,\lambda,q}|  \right\} \nonumber \\
            &+ \max\left\{ \mathcal{S}_b\left(f_L^{b,\lambda,q}\right), \mathcal{S}_b\left(f_{L-1}^{b,\lambda,q}\right) \right\} \nonumber \\
            &+ \mathcal{S}_\lambda\left( f_L^{b,\lambda,q}\right) +\mathcal{S}_q\left( f_L^{b,\lambda,q}\right) \,.
            \label{eq: error estimate}
\end{align}
and we select the parameters in such a way that this error estimate is
minimized. We also compare the sensitivities at different orders $L$ and  search
for a minimum in the dependence on $L$.

Explicitly, we scan the parameter space in the range
$
      (b,\lambda,q) \in [0,50] \times [0,2.5] \times [0, 0.5] 
$
in steps of $\delta_b = 0.5$, $\delta_\lambda = 0.02, \delta_q = 0.02$ and compute $E_{f,L}^{b,\lambda,q}$ for each parameter set.
We show the results of this scan for the inverse correlation length exponent $1/\nu$ in Fig.~\ref{fig: scan nuinv N8}.
The gobal minimum $\overline{E}$ of this scan marks the apparently best set of resummation parameters.
However, this minimum is not always sharp and there are other sets of parameters which are almost equally likely.
Therefore, we compute a weighted mean of all points which lie below a relative error of $E_i/\overline{E}<3$.
For the weights we use $w_i=1/E_i^2$ and plot the result as a dashed line in the parameter scans, see, e.g., in the right panel of Fig.~\ref{fig: scan nuinv N8}.
From the error estimate at the minimum we compile the error for the resummation as $3\overline{E}$.
The resulting error bars are consistent with the spread of the critical exponents around the optimum.
We provide more detail and numerical data on the error estimates in App.~\ref{app:error}.

\subsection{Graphene case ($N=8$)}

One of the quantum critical points that can be described by the Gross-Neveu-Yukawa model with $N=8$ is the semimetal-CDW transition of spin-1/2 electrons in graphene~\cite{PhysRevLett.97.146401}.
With the Borel resummation described in Sec.~\ref{sec:Borelresummation} applied to the $4-\epsilon$ expansion of the GNY model, we obtain the critical exponents $1/\nu\approx 0.993(27)$, $\eta_\phi\approx 0.704(15)$ and $\eta_\psi\approx 0.043(12)$.
This compares very well with the perturbative RG estimates which we obtained using the combination of $2+\epsilon$ and $4-\epsilon$ expansions in terms of the two-sided Pad\'e approximants, Sec.~\ref{sec:Pade}, and the polynomial interpolation Sec.~\ref{sec:polyinterpolation}. In particular, for the fermion anomalous dimension, the agreement is remarkable since, in this case, the one-sided Pad\'e estimates around each critical dimension, individually, seems to significantly overestimate the value at $D=3$ even at order $\mathcal{O}(\epsilon^4)$~\cite{Gracey:2016mio,PhysRevD.96.096010}.
Some deviation between our different resummation approaches at the $\lesssim 5\%$ level can be observed for the boson anomalous dimension, though.
%
\begin{table}[t]
      \caption{\textit{Graphene case at $N =8$}: Comparison of the resummed critical exponents obtained from the resummation algorithm (Sec. V) with the results from the polynomial interpolation (Sec. IV), the two-sided Pad\'e approximations (Sec. III) with with results from other theoretical methods.}
        \label{tab: graphene case N8}
        \begin{tabular*}{\linewidth}{@{\extracolsep\fill}llll}
             \hline\hline
             $N = 8$                               & $\nu^{-1}$          & $\eta_\phi$           & $\eta_\psi$ \\
             \hline
             Sec. \textbf{\ref{sec:Pade}}  & 1.004               & 0.735                 & 0.042       \\
             Sec. \textbf{\ref{sec:polyinterpolation}}    & 0.982               & 0.731                 & 0.043       \\
             Sec. \textbf{\ref{sec:Borelresummation}}      & 0.993(27)           & 0.704(15)             & 0.043(12)   \\
             \hline
             large-$N$\cite{Gracey:1992cp,Gracey:1993kc,Gracey:1993kb,Karkkainen:1993ef}     & 0.952               & 0.743                 & 0.044       \\
             conformal bootstrap\cite{Iliesiu:2017nrv}             & 0.88                & 0.742                 & 0.044       \\
             functional RG\cite{PhysRevB.94.245102}& 0.994(2)            & 0.7765                & 0.0276      \\
             Monte Carlo\cite{Chandrasekharan:2013aya}  & 1.20(1)             & 0.62(1)               & 0.38(1)     \\
             Monte Carlo\cite{Karkkainen:1993ef}   & 1.00(4)             & 0.754(8)              & --          \\
             Monte Carlo\cite{Schmidt:2017}        & 1.07(4)             & --                    & --          \\
             \hline\hline
        \end{tabular*}
\end{table}
%
We have compiled our results for $N=8$ from the two-sided Pad\'e approximants, the polynomial interpolation and the Borel resummation in Tab.~\ref{tab: graphene case N8}.
Here, to summarize all of the considerations from Secs.~\ref{sec:Pade},~\ref{sec:polyinterpolation} and \ref{sec:Borelresummation}, we build the simple average of the estimated critical exponents and transfer the error  from the Borel resummation
\begin{align}
 \frac{1}{\nu}\approx 0.99(3)\,,\quad \eta_\phi\approx 0.72(2)\,,\quad \eta_\psi\approx 0.043(1)\,.
\end{align}
In Tab.~\ref{tab: graphene case N8}, we also compare to other methods and refer to the more detailed discussions of these results in Secs.~\ref{sec:Pade} and~\ref{sec:polyinterpolation}.
We would like to emphasize, again, the excellent agreement of our results for the boson and fermion anomalous dimensions with the conformal bootstrap estimates~\cite{Iliesiu:2017nrv}.
The superior agreement between two independent theoretical approaches builds a strong case for these values.
At the same time, the available quantum Monte Carlo results for the anomalous
dimensions at $N=8$~\cite{Chandrasekharan:2013aya} show significant deviations
from the pRG and conformal bootstrap results, suggesting that the QMC approach
might still be affected by the finite system sizes.
Surprisingly, the earliest numerical lattice studies of the chiral Ising universality class at $N=8$~\cite{Karkkainen:1993ef} agree best with our results for the boson anomalous dimension.

We further note that there is still a sizable difference in the estimates of the inverse correlation length exponent between the different approaches which needs to be resolved in future studies.
In particular, it would be very interesting to obtain conformal bootstrap estimates of the correlations length exponent directly from universal bounds.

\subsection{Spinless fermions on the honeycomb ($N=4$)}

\begin{table}[t]
     \caption{\textit{Spinless honeycomb fermions at $N =4$}: Comparison of the resummed critical exponents obtained from the resummation algorithm (Sec. V) with the results from the polynomial interpolation (Sec. IV), the two-sided Pad\'e approximations (Sec. III) with results from other theoretical methods.}
        \label{tab: spinless fermions N4}
        \begin{tabular*}{\linewidth}{@{\extracolsep\fill}llll}
             \hline\hline
             $N = 4$                               & $\nu^{-1}$          & $\eta_\phi$           & $\eta_\psi$ \\
             \hline
             Sec. \textbf{\ref{sec:Pade}}   & \textcolor{gray}{0.961}               & \textcolor{gray}{0.480}                 & \textcolor{gray}{0.086}       \\
             Sec. \textbf{\ref{sec:polyinterpolation}}  & \textcolor{gray}{1.040}               & \textcolor{gray}{0.397}                 & \textcolor{gray}{0.140}       \\
             Sec. \textbf{\ref{sec:Borelresummation}}   & 1.114(33)           & 0.487(12)             & 0.102(12)   \\
             \hline
             large-$N$\cite{Gracey:1992cp,Gracey:1993kc,Gracey:1993kb}     & 0.938               & 0.509                 & 0.1056      \\
             conformal bootstrap\cite{Iliesiu:2017nrv}             & 0.76                & 0.544                 & 0.084       \\
             functional RG\cite{PhysRevB.94.245102}& 1.075(4)            & 0.5506                & 0.0654      \\
             Monte Carlo\cite{PhysRevD.96.114502}  &  1.14(2)      & 0.54(6)              & --          \\
             Monte Carlo\cite{Schmidt:2017}        & 1.096(34)           & --                    & --          \\
             Monte Carlo\cite{1367-2630-17-8-085003} & 1.30(5)             & 0.45(2)               & --           \\
             \hline\hline
        \end{tabular*}
\end{table}

The universality class of the semimetal-CDW transition of spinless fermions on the honeycomb lattice is described by the $N=4$ GN model~\cite{PhysRevLett.97.146401,PhysRevB.79.085116}. Due to its simplicity and paradigmatic role, it has been subject to many studies, see Ref.~\cite{0953-8984-29-4-043002} for a recent review.
Within our pRG approach we have found that the two-sided Pad\'e approximants and the polynomial interpolation presented in Secs.~\ref{sec:Pade} and~\ref{sec:polyinterpolation}, respectively, are problematic, because the critical exponents from the $2+\epsilon$ expansion exhibit a pole at $N=2$.
Therefore, the dimensional interpolation in $D\in [2,4]$ breaks down in the vicinity of $N=2$ and we find that this is already shows at $N=4$.
For completeness, however, we also show the values for the critical exponents that we obtain from these two interpolations in Tab.~\ref{tab: spinless fermions N4} in gray fonts.
The Borel resummation from Sec.~\ref{sec:Borelresummation} exclusively uses the $4-\epsilon$ expansion which is not plagued by this pole structure and, here, we obtain $1/\nu\approx 1.114(33)$, $\eta_\phi\approx 0.487(12)$ and $\eta_\psi\approx 0.102(12)$.

For $N=4$, the agreement of the boson and fermion anomalous dimensions with the conformal bootstrap estimates from Ref.~\cite{Iliesiu:2017nrv} is not as good as in the case $N=8$, where we also had more reliable results based on the dimensional interpolation in $D\in [2,4]$.
Still, within the error bars extracted from the resummation procedure, the estimates deviate only on the $\sim 10\%$ level.
We would like to mention that for the anomalous dimensions, the case $N=4$ turns out to show the largest deviation from the conformal bootstrap results.
For other N, i.e. $N=1,2$ and $N\gtrsim 6$ the agreement between these two independent methods is much better, see below.
It will be interesting to see, whether this can be fully resolved by going to even higher loop orders.
We would also like to mention that there is a series of Monte Carlo estimates for the boson anomalous dimension, i.e. $\eta_\phi=0.303(7)$~\cite{1367-2630-16-10-103008}, $\eta_\phi=0.45(2)$~\cite{1367-2630-17-8-085003}, $\eta_\phi=0.275(25)$~\cite{PhysRevB.93.155157}, and $\eta_\phi=0.54(6)$~\cite{PhysRevD.96.114502}.
The different results have been achieved using various Monte Carlo methods and different system sizes.
The last result from Ref.~\cite{PhysRevD.96.114502} was obtained with the largest system size. It seems to fit very well with the conformal bootstrap result and within the~$\sim 10\%$ range also with our estimate.
In view of the better control, we have for $N=8$ it would therefore be very interesting to explore this case with the same system sizes, too, and to have improved data for the fermion anomalous dimension for all cases.

The correlation-length exponent from the conformal bootstrap,
$1/\nu=0.76$~\cite{Iliesiu:2017nrv}, on the other hand, is very far away from
all the other approaches, in particular, also from the recent Monte Carlo
simulations, $1/\nu=1.14(2)$~\cite{PhysRevD.96.114502} which agreed well for
the anomalous dimension with the Borel resummed estimates from Sec.~\ref{sec:Borelresummation}.
Here, we note that our Borel resummed pRG estimate of $1/\nu=1.114(33)$ agrees very well with the Monte Carlo result from Ref.~\cite{PhysRevD.96.114502} as well as with another recent Monte Carlo estimate~\cite{Schmidt:2017} where $1/\nu=1.096(34)$. Unfortunately, the latter study does not provide data for the anomalous dimensions.

\begin{table}[t]
      \caption{\textit{Emergent SUSY at $N =1$}: Comparison of the resummed critical exponents obtained from the resummation algorithm (Sec. V) with the results from complementary methods:
                    Functional Renormalization Group (FRG) and conformal bootstrap.
                    Here, we have determined the conformal bootstrap estimate of $1/\nu$  from the scaling relation Eq.~\eqref{eq: super scaling}.}
        \label{tab: SUSY N1}
        \begin{tabular*}{\linewidth}{@{\extracolsep\fill}llll}
             \hline\hline
             $N = 1$                               & $\nu^{-1}$          & $\eta_\phi$           & $\eta_\psi$ \\
             \hline
             Sec. \textbf{\ref{sec:Borelresummation}}    & 1.415(12)           & 0.1673(27)             & 0.1673(27)   \\
             conformal bootstrap\cite{Iliesiu:2017nrv}   & {\sl 1.418}                & 0.164                 & 0.164       \\
             functional RG\cite{Gies2017}       & 1.395            & 0.167                & 0.167      \\
             \hline\hline
        \end{tabular*}
\end{table}

\subsection{Emergent supersymmetry ($N=1$)}

The emergent supersymmetry scenario~\cite{Grover280,Fei2016} for the case~$N=1$ features a supersymmetric scaling relation~\cite{Gies:2009az,Heilmann:2014iga} which connects the inverse correlation length exponent, dimensionality and the anomalous dimension
\begin{align}
      \nu^{-1}=(D-\eta)/2\,. \label{eq: super scaling}
\end{align}
Note, that the anomalous dimensions of the boson and the fermion are equal in that case $\eta = \eta_\psi = \eta_\phi$.
Eq.~\eqref{eq: super scaling} therefore allows for a number of non-trivial checks for the estimates of the critical exponents.
In  previous works, cf. Refs.~\cite{Mihaila:2017ble,PhysRevD.96.096010},
it has been shown that Eq.~\eqref{eq: super scaling} holds exactly order
by order in the epsilon expansion up to $\mathcal{O}(\epsilon^4)$. Moreover,
the estimates for all critical exponents from simple Pad\'e approximants
already showed very good agreement with the conformal
bootstrap~\cite{Iliesiu:2017nrv} and the functional RG
results~\cite{Gies2017}.
Borel resummation yields the values $1/\nu\approx 1.415(12)$, $\eta_\phi\approx 0.1673(27)$ and $\eta_\psi\approx 0.1673(27)$.
Since the anomalous dimensions of the boson and fermion coincide order by order in the
epsilon expansion, also the Borel resummation provides the same values for both exponents.
Further, within the error bars, the supersymmetric scaling relation,  Eq.~\eqref{eq: super scaling}, is fulfilled as expected.
Here, we obtain excellent agreement among all available theoretical approaches, see Tab.~\ref{tab: SUSY N1}.
Using the scaling relation in Eq.~\eqref{eq: super scaling}, we show that this agreement even holds for the inverse correlation length which, for other $N$, shows clear deviations.
We also would like to mention that our results continuously connect to the limit $N\to 0$ where we recover critical exponents compatible with the three-dimensional Ising model.
We show this in Fig.~\ref{fig:criticalexponentsN} together with the data point from the highly accurate conformal bootstrap results~\cite{Kos:2016ysd}.

\subsection{Other cases ($N=2$ and $N=20$)}

\begin{figure*}[t]
    \includegraphics[scale=1.0]{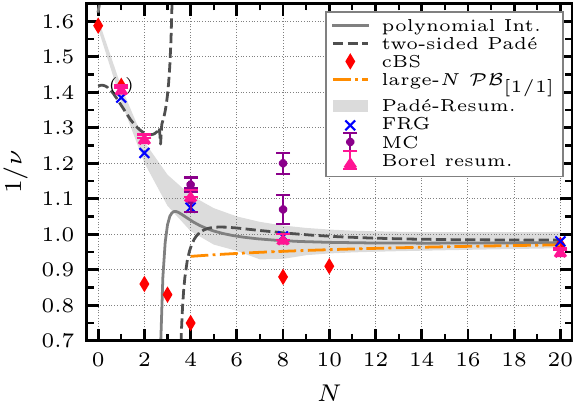}
    \includegraphics[scale=1.0]{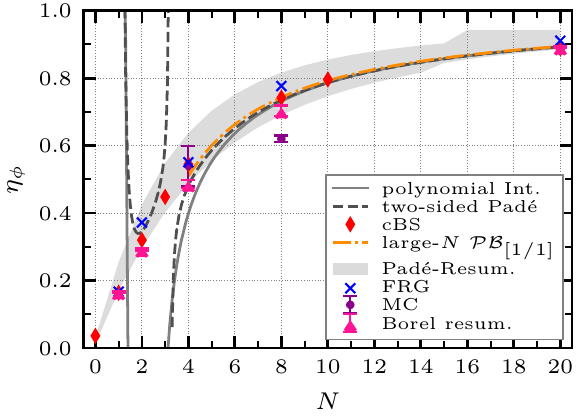}
    \includegraphics[scale=1.0]{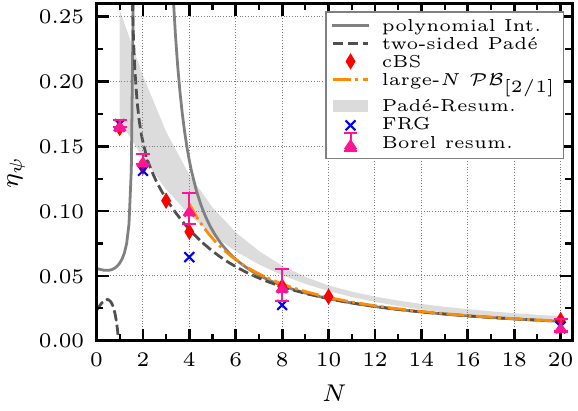}
    \caption{\textit{Chiral Ising universality} in $D=3$: Overview plot of the three examined critical exponents as the correlation-length exponent $\nu^{-1}$ (left panel), the boson anomalous
     dimension $\eta_\phi$ (medium panel) and the fermion anomalous dimension $\eta_\psi$ (right panel) for different spinor component numbers $N\in[1,20]$.
     For comparison, the values from Monte Carlo (MC) calculations and the functional renormalization group (FRG), as well as the conformal bootstrap (cBS) were also plotted.
     On top of that we applied a Pad\'e-Borel resummation on the large-$N$
     expansions as in Ref.~\cite{Karkkainen:1993ef}.}
    \label{fig:criticalexponentsNallmethods}
\end{figure*}

\begin{table}[t]
     \caption{\textit{Chiral universality at $N =2$}: Comparison of the resummed critical exponents obtained from the resummation algorithm (Sec. V) with the results from complementary methods:
                   Functional Renormalization Group (FRG) and conformal bootstrap (cBS).}
                    \label{tab:N2}
        \begin{tabular*}{\linewidth}{@{\extracolsep\fill}llll}
             \hline\hline
             $N = 2$                               & $\nu^{-1}$          & $\eta_\phi$           & $\eta_\psi$ \\
             \hline
             Sec. \textbf{\ref{sec:Borelresummation}}  & 1.276(15)           & 0.2934(42)             & 0.1400(39)   \\
             conformal bootstrap\cite{Iliesiu:2017nrv}   &  0.86                &  0.320                 & 0.134      \\
             functional RG\cite{Vacca2015}            & 1.229                & 0.372                    & 0.131 \\
             \hline\hline
        \end{tabular*}
\end{table}

In addition to these cases, we compare the two additional choices $N=2$ and $N=20$.
At $N=2$ no interpolation to the Gross-Neveu model in the $2+\epsilon$ expansion is possible due to the pole in the critical exponents, cf. App.~\ref{app:JohnGN}.
Other methods which provide results in this case are the conformal bootstrap and the functional RG and we compile these results for comparison in Tab.~\ref{tab:N2}.
This fits into the general picture that our anomalous dimension agree very well with the conformal bootstrap results, but the inverse correlation length agrees better with the functional RG.

\begin{table}[b]
     \caption{\textit{Chiral universality at $N =20$}: Comparison of the resummed critical exponents obtained from the resummation algorithm (Sec. V) with the results from the polynomial interpolation (Sec. III), the two-sided Pad\'e approximations (Sec. IV) and complementary methods:
                   Functional Renormalization Group (FRG), conformal bootstrap (cBS), Monte Carlo (MC) and large $N$ calculations.}
                      \label{tab:N20}
        \begin{tabular*}{\linewidth}{@{\extracolsep\fill}llll}
             \hline\hline
             $N = 20$                               & $\nu^{-1}$          & $\eta_\phi$           & $\eta_\psi$ \\
             \hline
             Sec. \textbf{\ref{sec:Pade}}    & 0.9840               & 0.893                 & 0.0151       \\
             Sec. \textbf{\ref{sec:polyinterpolation}}   & 0.9763               & 0.893                 & 0.0150       \\
             Sec. \textbf{\ref{sec:Borelresummation}}   & 0.9580(75)           & 0.893(9)             & 0.0120(48)   \\
             \hline
             large-$N$\cite{Gracey:1992cp,Gracey:1993kc,Gracey:1993kb}    & 0.970                & 0.894                  & 0.0152 \\
             conformal bootstrap\cite{Iliesiu:2017nrv}   &  0.97                &  0.888                 & 0.016      \\
             FRG, App.~\ref{app:FRG}                        & 0.980                 & 0.911                & 0.011\\
             \hline\hline
        \end{tabular*}
\end{table}

Finally, we study the case $N=20$, cf.~Tab.~\ref{tab:N20}, which should already be well located in the large-$N$ regime. It is accessible by all the resummation approaches we introduced in this paper and shows very good agreement between the two-sided Pad\'e approximants, the polynomial interpolation and the Borel resummation. Moreover, it agrees excellently with the direct large-$N$ results~\cite{Karkkainen:1993ef}, the conformal bootstrap~\cite{Iliesiu:2017nrv} and also very well with functional RG, which we have carried out in a simple approximation by ourselves see App.~\ref{app:FRG}.
We expect that this level of agreement generally holds in the large-$N$ regime.
This could therefore be a very good benchmark case for upcoming large-scale Monte Carlo calculations as the large-$N$ limit reduces uncertainties within the approximations required for the renormalization group methods.
We have compiled an overview plot with all of our data for $D=3$ and comparisons for the range $N\in [0,20]$ in Fig.~\ref{fig:criticalexponentsNallmethods} which supports this expectation.

\section{Conclusions}\label{sec:conculsions}

We studied the universality classes of chiral Ising or Gross-Neveu(-Yukawa)
models using the perturbative renormalization group up to four-loop order.
Employing various resummation and interpolation techniques, in particular, dimensional interpolations between two and four dimensions as well as Borel resummation, we calculated estimates for the critical exponents.

To describe the physically interesting case of interacting electrons on graphene's honeycomb lattice which undergo a quantum phase transition towards an ordered charge density wave state, we have focused on the $N=8$ Gross-Neveu and Gross-Neveu-Yukawa models.
We have found that all of our approaches to extract critical exponents in $D=2+1$ within the perturbative RG, i.e. the two-sided Pad\'e approximants, the polynomial interpolation and also the Borel resummation, converge order by order towards a stable and compatible set of values, as compiled in Tab.~\ref{tab: graphene case N8}.
This is true for all the exponents that we have studied, i.e. the inverse correlation length exponent and both anomalous dimensions.

While resummation techniques for this type of theories is little explored and we cannot exclude unexpected behavior at higher orders in the loop expansion, the stability of our results still suggests they could be considered as very reasonable estimates.
Moreover, for the boson and fermion anomalous dimensions at $N=8$, we find excellent agreement of our results with the universal bounds provided by the conformal bootstrap method~\cite{Iliesiu:2017nrv}, which builds a strong case from independent theoretical methods for the validity of these values.
The corresponding values from Quantum Monte Carlo~\cite{Chandrasekharan:2013aya} do not agree for the anomalous dimensions and it will be very interesting to see whether this issue can be resolved in the future, e.g., by increasing system sizes.
In particular, the fermion anomalous dimension is off by about an order of magnitude in the QMC simulations~\cite{Chandrasekharan:2013aya}.
Reconciliation with the field theoretical results may be achieved by going to larger lattice sizes as suggested from similar calculations for the case $N=4$~\cite{PhysRevD.96.114502}.
We also note that there is still a significant difference in the estimates for the inverse correlation length exponent between the perturbative RG and the conformal bootstrap.
This difference is most pronounced for a range in the number of spinor components, $2\leq N\leq 8$, while for larger $N$ it disappears and all the exponents agree very well.
Within the conformal bootstrap~\cite{Iliesiu:2017nrv}, in contrast to the anomalous dimensions, the correlation length exponent estimate is not obtained from universal bounds but from the extremal functional approach.
It would therefore be very interesting to have conformal bootstrap estimates of the correlations length exponent directly from universal bounds, too.
We have also found very good agreement with the conformal bootstrap and the functional RG approach for the case of emergent supersymmetry, $N=1$. Here, deviations across the different approaches are only found on the level of below 2\%.

For the future, we identify two main directions for this line of research. Firstly, to fully establish convergence of the critical exponents within the perturbative RG approaches, loop calculations beyond fourth order are required. In view of the recent developments in computational technology and the mathematical insights into the structure of Feynman diagrams, this may be challenging but possible.
Secondly, there are a number of very interesting quantum critical points of interacting Dirac fermions, which are not captured by the chiral Ising universality class.
For example, superconducting or magnetic transitions of Dirac fermions exhibit critical behavior which are described by a coupling to $U(1)$ or a $O(3)$ order parameters, respectively.
The corresponding universality classes are known as the chiral XY and chiral Heisenberg universality classes~\cite{Rosenstein:1993zf} and in perturbative RG are also known for higher orders~\cite{PhysRevD.96.096010,Gracey:2018qba}.
Moreover, for these and even more exotic transitions, there is recent QMC data~\cite{PhysRevB.91.165108,PhysRevX.6.011029,2015arXiv151207908L,Li:2017dkj,Otsuka:2018kcb,2018arXiv180500478X,PhysRevB.95.241103}, which would be interesting to compare to.
It can also be expected that more conformal bootstrap results will be available, soon, see,  e.g., Ref.~\cite{Bobev:2015vsa}.
As has been the case for the chiral Ising universality class, the critical exponents of the chiral XY and chiral Heisenberg universality classes, have not been satisfactorily settled, yet.
For these models, a thorough study of the critical exponents from resummation and interpolation techniques is under way.

An alternative and promising approach for the quantitative characterization of the 2+1 dimensional chiral universality classes could also be the analysis of low-energy finite-size torus spectra at  quantum critical points, cf.~Refs.~\cite{PhysRevLett.117.210401,PhysRevB.96.035142}.

Moreover, we note that it has recently been possible to observe strongly-correlated behavior and superconductivity in graphene-based systems, i.e. in twisted bilayer graphene~\cite{Jarillo-Herrero:2018a,Jarillo-Herrero:2018}.
While it is currently not clear which mechanism underlies these transitions, the chiral universality classes may become relevant in this context, too.
This, however, is still subject to discussion~\cite{XuBalents2018,2018arXiv180309699Y,2018arXiv180309742P,2018arXiv180400159G,2018arXiv180403162D} and further experimental data and theoretical studies are required to settle the situation.

\bigskip
\paragraph*{Acknowledgments}
The authors are grateful to Holger Gies, John A. Gracey, Igor F. Herbut, Benjamin Knorr, Zohar Komargodski, Zi-Yang Meng, Anders Sandvik, Simon Trebst, Andreas Wipf and Hong Yao for discussions and David Poland for correspondence. We thank Sebastian Diehl, John A. Gracey and Stefan Wessel for comments on the manuscript. MMS and BI were supported by Deutsche Forschungsgemeinschaft through the Collaborative Research Center SFB~1238, TP~C04. LNM was supported by Deutsche
Forschungsgemeinschaft (Contract MI 1358, Heisenberg
program).

\appendix

\section{GNY model beta and gamma functions}\label{app:betagamma}

For the convenience of the reader, we give below the analytic expressions for the RG functions that we used in the numerical analysis.
Here, and in the following, we use $N_\text{f}=N/d_\gamma$ to display the RG functions.
For the Ising GNY model, the beta function contributions to the Yukawa coupling explicitly read~\cite{PhysRevD.96.096010}
\begin{align}
	\beta_{y}^{\text{(1L)}}&=(3+2N_{\mathrm{f}})y^2\,,\\[5pt]
	\beta_{y}^{\text{(2L)}}&=24y\lambda(\lambda- y)-\big(\frac{9}{8}+6N_{\mathrm{f}}\big)y^3\,,\\[5pt]
	\beta_{y}^{\text{(3L)}}&=\frac{y}{64}\Big(1152 (7+5N_{\mathrm{f}}) y^2\lambda+192 (91-30 N_{\mathrm{f}}) y \lambda ^2\nonumber\\
	&+\big(912 \zeta_3-697+2 N_{\mathrm{f}} (67+112 N_{\mathrm{f}}+432 \zeta_3)\big)y^3\nonumber\\
	&-13824 \lambda ^3\Big)\,.
\end{align}
Here $\zeta_z$ is the Riemann zeta function. The four-loop contribution is listed below, together with all the other four-loop contributions.
Accordingly, the beta function for the quartic scalar coupling are composed of
\begin{align}
	\beta_{\lambda}^{\text{(1L)}}&=36 \lambda ^2+4N_{\mathrm{f}} y \lambda -N_{\mathrm{f}} y^2\,,\\
	\beta_{\lambda}^{\text{(2L)}}&=4 N_{\mathrm{f}} y^3+7N_{\mathrm{f}} y^2 \lambda-72N_{\mathrm{f}} y \lambda ^2-816 \lambda ^3\,,\\[5pt]
	\beta_{\lambda}^{\text{(3L)}}&=\frac{1}{32} \Big(6912 (145+96 \zeta_3) \lambda ^4+49536 N_{\mathrm{f}} y \lambda ^3\nonumber\\[5pt]
	&-48 N_{\mathrm{f}} (72 N_{\mathrm{f}}-361-648 \zeta_3)y^2\lambda ^2\nonumber\\
	&+2 N_{\mathrm{f}} (1736 N_{\mathrm{f}}-4395-1872 \zeta_3)y^3 \lambda\nonumber\\
	&+N_{\mathrm{f}} (5-628 N_{\mathrm{f}}-384 \zeta_3)y^4\Big)\,.
\end{align}
For the contributions to the RG gamma function corresponding to wave function renormalization of the fermion derivative term, we find
\begin{align}
	\gamma_{\psi}^{\text{(1L)}}&=\frac{y}{2}\,,\\[5pt]
	\gamma_{\psi}^{\text{(2L)}}&=-\frac{y^2}{16} (12 N_{\mathrm{f}}+1)\,,\\[5pt]
	\gamma_{\psi}^{\text{(3L)}}&=\frac{y^3}{128}\left(48 \zeta _3+4 N_{\mathrm{f}} (47-12 N_{\mathrm{f}})-15\right)\nonumber\\
	&\quad+6 \lambda  y^2-\frac{33 \lambda ^2 y}{2}\,.
\end{align}
The gamma function corresponding to the wave function renormalization of the derivative term of the scalar order parameter reads
\begin{align}
	\gamma_{\phi}^{\text{(1L)}}&=2N_{\mathrm{f}}y\,,\\
	\gamma_{\phi}^{\text{(2L)}}&=24 \lambda ^2-\frac{5 N_{\mathrm{f}} y^2}{2}\,,\\
	\gamma_{\phi}^{\text{(3L)}}&=-216 \lambda ^3+\frac{1}{32} N_{\mathrm{f}} y^3 \left(48 \zeta _3+200 N_{\mathrm{f}}+21\right)\nonumber\\
	&\quad+30 \lambda  N_{\mathrm{f}} y^2-90 \lambda ^2 N_{\mathrm{f}} y\,.
\end{align}
Finally, the scaling of the quadratic scalar operator is given by the following contributions to the RG gamma function $\gamma_{\phi^2}$,
\begin{align}
	\gamma_{\phi^2}^{\text{(1L)}}&=-12\lambda\,,\\
	\gamma_{\phi^2}^{\text{(2L)}}&=144 \lambda ^2-2 N_{\mathrm{f}} y (y-12 \lambda )\,,\\
	\gamma_{\phi^2}^{\text{(3L)}}&=\frac{3}{2} N_{\mathrm{f}}y^2\lambda  \left(24 N_{\mathrm{f}}-120 \zeta _3-11\right)-6264 \lambda ^3\nonumber\\
	&\quad-4 N_{\mathrm{f}} y^3 \left(4 N_{\mathrm{f}}-9+3 \zeta _3\right)-288 N_{\mathrm{f}} y \lambda ^2\,.
\end{align}
To complete the set of RG beta and gamma functions at the available order, we now also display the four-loop contributions,
\begin{widetext}
\begin{align}\label{eq:beta4lci}
	\beta_{y}^{\text{(4L)}}&=-\frac{5}{2} \zeta _5 (42 N_{\mathrm{f}}+43) y^5+\frac{\left(32 \pi ^4 (2 N_{\mathrm{f}}+3) (18 N_{\mathrm{f}}+19)+40 N_{\mathrm{f}} (8 N_{\mathrm{f}} (44 N_{\mathrm{f}}-899)+29721)+457935\right) y^5}{7680}\\
	&+\frac{\lambda}{8} (8 N_{\mathrm{f}} (12 N_{\mathrm{f}}-683)-2829) y^4-\frac{1}{2} \lambda ^2 (4 N_{\mathrm{f}} (6 N_{\mathrm{f}}+635)+4455) y^3+36 \lambda ^3 (8 N_{\mathrm{f}}-455) y^2\nonumber\\
	&-\frac{1}{8} \zeta _3 y^2 \left(-41472 \lambda ^3+(4 N_{\mathrm{f}} (125 N_{\mathrm{f}}+331)-5)
        y^3+432 \lambda  (12 N_{\mathrm{f}}+7) y^2-864 \lambda ^2 (6 N_{\mathrm{f}}-25) y\right)+14040
        \lambda ^4 y\,,\nonumber\\
& + \Delta_3 N_{\mathrm{f}} (1 + 107 \zeta _3 - 125  \zeta _5) y^5\,,\nonumber\\[10pt]
	\beta_{\lambda}^{\text{(4L)}}&=41472 \left(-39 \zeta _3-60 \zeta _5+\frac{\pi ^4}{10}-\frac{3499}{96}\right) \lambda ^5+\frac{1}{240} \lambda  N_{\mathrm{f}} y^4 \left(-60 \zeta _3 \left(912 N_{\mathrm{f}}^2-4156 N_{\mathrm{f}}-4677\right)\right.\\
	&\left.+1200 \zeta _5 (157-168 N_{\mathrm{f}})-4 \pi ^4 (450 N_{\mathrm{f}}+41)+25 (4 N_{\mathrm{f}} (337 N_{\mathrm{f}}+3461)+5847)\right)\nonumber\\
	&+\frac{N_{\mathrm{f}} y^5 \left(480 \zeta _3 (12 N_{\mathrm{f}} (14 N_{\mathrm{f}}-15)+277)+2400 \zeta _5 (128 N_{\mathrm{f}}+65)+8 \pi ^4 (64 N_{\mathrm{f}}-77)+160 N_{\mathrm{f}} (1289-386 N_{\mathrm{f}})-67095\right)}{1920}\nonumber\\
	&+\frac{1}{80} \lambda ^2 N_{\mathrm{f}} y^3 \left(835200 \zeta _5+1920 \zeta _3 (3 N_{\mathrm{f}} (4 N_{\mathrm{f}}-61)+19)+72 \pi ^4 (24 N_{\mathrm{f}}+31)-40 N_{\mathrm{f}} (288 N_{\mathrm{f}}+15649)+1057825\right)\nonumber\\
	&+\frac{4}{5} \lambda ^3 N_{\mathrm{f}} y^2 \left(-86400 \zeta _5+540 \zeta _3 (4
        N_{\mathrm{f}}-69)+7890 N_{\mathrm{f}}-288 \pi ^4-72605\right)+\frac{36}{5} \left(-17280
        \zeta_3+96 \pi ^4-6775\right) \lambda ^4 N_{\mathrm{f}} y\,.\nonumber
\end{align}
The symbol $\Delta_3$ should be set to $\Delta_3=1$, see Ref.~\cite{Mihaila:2017ble} for more details on the dimensional regularization scheme.
The four-loop contributions to the gamma functions read
\begin{align}
	\gamma_{\psi}^{\text{(4L)}}&=\frac{y}{393216} \Bigg(134479872 \lambda ^3+y^3 \left(-884736 \zeta _3-5 \left(384 \left(256 \zeta _5-893\right)+\frac{377339 \pi ^4}{90}\right)\right.\\
	&\left.+16 N_{\mathrm{f}} \left(-164352 \zeta _3+N_{\mathrm{f}} \left(1536 \left(16 \zeta _3-3\right) N_{\mathrm{f}}-74752\right)-\frac{1536 \pi ^4}{5}+53440\right)+\frac{303611 \pi ^4}{18}\right)\nonumber\\
	&-288 \lambda  y^2 \left(512 \left(93-32 \zeta _3\right)+8 \left(7424+\frac{927 \pi ^4}{4}\right) N_{\mathrm{f}}-\frac{1}{18} \pi ^4 (33372 N_{\mathrm{f}}+7079)+\frac{7079 \pi ^4}{18}\right)\nonumber\\
	&+96 \lambda ^2 y \left(221184 \zeta _3+344064 N_{\mathrm{f}}-656384\right)\Bigg)\,,\nonumber\\[10pt]
	\gamma_{\phi}^{\text{(4L)}}&=14040 \lambda ^4+\frac{1}{256} \lambda  N_{\mathrm{f}} y^3 \left(768 \left(16 \zeta _3-83\right)-19456 N_{\mathrm{f}}\right)+\frac{1}{32} \lambda ^2 N_{\mathrm{f}} y^2 \left(256 \left(81 \zeta _3-91\right)-384 N_{\mathrm{f}}\right)+288 \lambda ^3 N_{\mathrm{f}} y\\
	&-\frac{N_{\mathrm{f}} y^4 \left(377856 \zeta _3+15360 \left(8 \zeta _5-29\right)+4 N_{\mathrm{f}} \left(162816 \zeta _3+256 \left(144 \zeta _3-101\right) N_{\mathrm{f}}+\frac{1536 \pi ^4}{5}-54016\right)+1024 \pi ^4\right)}{24576}\,,\nonumber\\[10pt]
	\gamma_{\phi^2}^{\text{(4L)}}&=1728 \left(18 \zeta _3+\frac{2 \pi ^4}{5}+187\right) \lambda ^4+\frac{3}{2} \lambda ^2 N_{\mathrm{f}} y^2 \left(5760 \zeta _3+4 \left(-48 \zeta _3-176\right) N_{\mathrm{f}}+\frac{48 \pi ^4}{5}+3796\right)\\
	&+\frac{1}{64} N_{\mathrm{f}} y^4 \left(-5376 \zeta _3+10080 \zeta _5+2 N_{\mathrm{f}} \left(320 \zeta _3+4480 \zeta _5+64 \left(18 \zeta _3-11\right) N_{\mathrm{f}}+\frac{48 \pi ^4}{5}-5208\right)-\frac{224 \pi ^4}{5}-2846\right)\nonumber\\
	&-\frac{3}{16} \lambda  N_{\mathrm{f}} y^3 \left(-5120 \zeta _3-5760 \zeta _5+4 N_{\mathrm{f}} \left(-672 \zeta _3+64 \left(2 \zeta _3-1\right) N_{\mathrm{f}}+\frac{16 \pi ^4}{3}-1618\right)+\frac{184 \pi ^4}{5}+12989\right)\nonumber\\
	&+36 \left(96 \zeta _3+313\right) \lambda ^3 N_{\mathrm{f}} y\,.\nonumber
\end{align}

\end{widetext}

\section{GN critical exponents from Ref.~\cite{Gracey:2016mio}}\label{app:JohnGN}

From the four-loop RG beta and gamma functions of the GN model in Ref.~\onlinecite{Gracey:2016mio} we have extracted the critical exponents as a function of general $N$. They read
 \begin{align}
 	\frac{1}{\nu}&=\epsilon+\frac{1}{2-N}\epsilon ^2-\frac{(N-3) }{2 (N-2)^2}\epsilon ^3\notag\\
 	&\hspace{-0.2cm}+\frac{6 \zeta _3 (11 N-34)+(N-1) (N+12)}{4 (N-2)^3}\epsilon ^4+\mathcal{O}(\epsilon^5)\,,\\[5pt]
 	\eta_\phi&=2-\frac{N}{N-2}\epsilon+\frac{1-N}{(N-2)^2}\epsilon ^2+\frac{(N-1) N }{2 (N-2)^3}\epsilon ^3\notag\\
 	&\hspace{-0.2cm}+\frac{(N-1)\left(2 \zeta _3 (N (N+7)-42)-(N-9) N+5\right)}{4 (N-2)^4} \epsilon ^4\notag\\
 	&+\mathcal{O}(\epsilon^5)\,,\\[5pt]
 	\eta_\psi&=\frac{N-1}{2 (N-2)^2}\epsilon ^2-\frac{(N-6) (N-1)}{4 (N-2)^3}\epsilon ^3\notag\\
 	&+\frac{(N-1) ((N-11) N+25)}{8 (N-2)^4} \epsilon ^4+\mathcal{O}(\epsilon^5)\,.
 \end{align}
 We note that these critical exponents have a pole at $N=2$, which is due to a factor of $(N-2)$ appearing in each loop order of the RG functions, cf.~Ref.~\onlinecite{Gracey:2016mio}.

\section{Error estimate for resummation}\label{app:error}

As described in Sec.~\ref{sec:Borelresummation} for the Borel resummation we performed extensive scans of parameter space $(b,\lambda,q) \in [0,50] \times [0,2.5] \times [0, 0.5]$ in steps
of $\delta_b = 0.5$, $\delta_\lambda = 0.02$ and $\delta_q = 0.02$ and computed
for each point $(b,\lambda,q)$  the error estimate $E_{f,L}^{b,\lambda,q}$. In Tab.~\ref{tab: minima} we list the minima $\overline{E}$ of these scans
and their position in the parameter space.
\begin{table}[t!]
      \caption{In this table we compiled the positions of the minima $\overline{E} = E_4^{b,\lambda,q}$ in the Borel resummation. While the minimum has the value $\tilde{f}^{b,\lambda,q}$ we use the points with a
      relative error of $E/\overline{E} \le 3$ to compute the weighted mean $\overline{\tilde{f}^{b,\lambda,q}}$ (last column) as explained in the text.}
      \label{tab: minima}
        \begin{tabular*}{\linewidth}{@{\extracolsep\fill}lllllll}
             \hline\hline
             $N=1$                  & $b$               & $\lambda$             & $q$                &   $E_4^{b,\lambda,q}$     & $\tilde{f}^{b,\lambda,q}$  & $\overline{\tilde{f}^{b,\lambda,q}}$ \\
             \hline
             $\nu^{-1}$             & 5.0               &   0.76                &  0.06              &  0.004                    &  1.414                     & 1.415  \\
             $\eta_\phi$            & 16.0              &  1.30                  & 0.00                &  0.0009                   &  0.1669                    & 0.1673 \\
             $\eta_\psi$            & 16.0              & 1.30                  &  0.00              &  0.0009                    &   0.1669                        & 0.1673\\
             \hline
             $N=2$                  & $b$               & $\lambda$             & $q$                &   $E_4^{b,\lambda,q}$     & $\tilde{f}^{b,\lambda,q}$  & $\overline{\tilde{f}^{b,\lambda,q}}$ \\
             \hline
             $\nu^{-1}$             & 11.0              &   0.74                &  0.12              &  0.005                    &  1.274                        & 1.276  \\
             $\eta_\phi$            & 19.5              &  1.26                  & 0.02               &  0.0014                    &   0.2929                        & 0.2934 \\
             $\eta_\psi$            & 11.0              & 1.08                   &  0.00              &  0.0013                    &   0.1392                        & 0.1400 \\
             \hline
             $N=4$                  & $b$               & $\lambda$             & $q$                &   $E_4^{b,\lambda,q}$     & $\tilde{f}^{b,\lambda,q}$  & $\overline{\tilde{f}^{b,\lambda,q}}$ \\
             \hline
             $\nu^{-1}$             & 20.5              &   0.74                &  0.16              &  0.010                    &  1.112                        & 1.114  \\
             $\eta_\phi$            & 22.5              &  1.10                  & 0.12               &  0.003                    &   0.489                        & 0.488 \\
             $\eta_\psi$            & 11.0              & 1.90                   &  0.00              &  0.0029                    &   0.0957                        & 0.0998 \\
             \hline
             $N=8$                  & $b$               & $\lambda$             & $q$                &   $E_4^{b,\lambda,q}$     & $\tilde{f}^{b,\lambda,q}$ &  $\overline{\tilde{f}^{b,\lambda,q}}$  \\
             \hline
             $\nu^{-1}$             & 31.5              &   0.74                &  0.16              &  0.009                    & 0.994                     & 0.993 \\
             $\eta_\phi$            & 27.5              &  1.00                  & 0.18               &  0.005                    & 0.706                     & 0.704  \\
             $\eta_\psi$            &  17.5             & 1.86                  &  0.22              &  0.004                    & 0.042                     & 0.043 \\
             \hline
             $N=20$                  & $b$               & $\lambda$             & $q$                &   $E_4^{b,\lambda,q}$     & $\tilde{f}^{b,\lambda,q}$ &  $\overline{\tilde{f}^{b,\lambda,q}}$  \\
             \hline
             $\nu^{-1}$             & 20.5              &   0.76                &  0.10              &  0.0025                    & 0.9571                    & 0.9580 \\
             $\eta_\phi$            & 31.0              &  0.86                  & 0.18               &  0.003                    & 0.894                     & 0.893  \\
             $\eta_\psi$            &  28.5             & 1.88                  &  0.18              &  0.0016                    & 0.0111                     & 0.0120 \\
             \hline\hline
        \end{tabular*}
\end{table}
The estimates for the critical exponents are then computed from the weighted mean of
all points in parameter space which lie below a relative error of $E_i/\overline{E} < 3$, i.e.
\begin{align}
 \overline{f} = \frac{\sum_{i \in \{E_i/\overline{E} < 3\}} w_i f(b_i,\lambda_i,q_i)}{\sum_{i \in \{E_i/\overline{E} < 3\}} w_i}
\end{align}
where $w_i = 1/E_i^2$ denote the weights. Note also that for $\eta_\psi$ at $N=8$ we omitted all points with $\eta_\psi > 0.053$ from the
weighted mean in order to extract a meaningful unbiased result.
We observe that values of the Borel parameter $b$ and the homographic shift $q$ increase for intermediate $N$ while the adjusting large order parameter
$\lambda$ is rather stable. The error estimate also increases for intermediate $N$ and decreases for very large values again, but is smallest in the SUSY case.
For the latter the Borel resummation method is able to determine the value to two significant digits.

\section{FRG and derivative expansion}
\label{app:FRG}

The FRG is a generalization of Wilson~RG. It is based on adding a scale-dependent regulator contribution to the action, i.e. $S\mapsto S_k \equiv S + \int_p\frac{1}{2}\Phi(-p)R_{k}(p)\Phi(p)$.
Then, the modified action $S_k$ gives a scale-dependent partition function $\mathcal Z_k=\int_\Lambda \mathcal{D}\Phi\, \exp(-S_k[\Phi])$, where $\Phi$ is a collective field variable of a specific model.
The central object of the FRG is the effective average action $\Gamma_k$, which is the Legendre transform of $\ln \mathcal Z_k$. For a suitably chosen regulator $R_k$, $\Gamma_k$ interpolates between the bare action $S$ at the ultraviolet cutoff ($k = \Lambda$) and the quantum effective action $\Gamma$ in the infrared ($k \to 0$). Its RG flow is governed by the exact equation~\cite{Wetterich:1992yh,Berges:2000ew}
\begin{align}\label{eqn:Wetterich}
\partial_k\Gamma_k = \frac{1}{2}\text{STr}\left[(\Gamma_k^{(2)}+R_k)^{-1}\partial_k R_k\right]\,.
\end{align}
Here, $\Gamma_k^{(2)}$ is the functional Hessian of $\Gamma_k$ with respect to~$\Phi$.
To obtain an approximate solution of Eq.~\eqref{eqn:Wetterich} for the chiral Ising model, we make a derivative expansion of $\Gamma_k$ and truncate beyond the leading order,
\begin{align}\label{eq:trunc}
	\Gamma_k=\hspace{-0.1cm}\int_ {\tau, \vec{x}}\hspace{-0.1cm}(Z_{\psi}\bar\psi\slashed{\partial}\psi+g \phi\bar\psi\psi-\frac{Z_\phi}{2}\phi\partial^2\phi+U(\rho)).
\end{align}
This is local potential approximation (LPA') with the scale-dependent effective potential $U(\rho)$ which is a functional of the field invariant $\rho=\frac{1}{2}\phi^2$.
We have further introduced the scale-dependent wave-function renormalizations $Z_\psi$ and $Z_\phi$, which are related to the anomalous dimensions by $\eta_\Phi=-(\partial_t Z_\Phi)/Z_\Phi$, with $\Phi \in \{\psi,\phi\}$ and $\partial_t := k \partial_k$.
To determine the fixed-points and critical exponents within the FRG, we define dimensionless quantities, i.e. the dimensionless effective potential $u$ and the dimensionless Yukawa coupling $\hat g$. They are given by $u(\hat\rho)=k^{-D}U(\rho)$ and $\hat g^2=k^{D-4}g^2/(Z_{\phi}Z_{\psi}^2)$, with $\rho=k^{D-2}Z_{\phi}^{-1} \hat\rho$.

The flow equations for $u, \hat g^2$ and the anomalous dimensions $\eta_\phi$ and $\eta_\psi$ are obtained by substituting the ansatz for $\Gamma_k$ into Eq.~\eqref{eqn:Wetterich} and applying suitable projection prescriptions.
For the chiral Ising model the LPA' flow equations read~\cite{Rosa:2000ju,Hofling:2002hj,Janssen:2014gea} for the effective potential
\begin{align}
\partial_t u&=-D u +(D-2+\eta_\phi)\hat\rho u^{\prime}-4N v_D l^\mathrm F(2 \hat g^2 \hat\rho)\notag\\[5pt]
&\quad+2v_D \left[l^\mathrm{B}( u^{\prime}+2\hat\rho  u^{\prime\prime})\right]\,,\notag
\end{align}
and the Yukawa coupling,
\begin{align}
\partial_t \hat g^2=(D-4+\eta_\phi+2\eta_\psi)\hat g^2+8v_{D}\hat g^4l_{11}^\mathrm{FB}(2\hat g^2\hat\rho,u^{\prime})\,.\notag
\end{align}
We have used $v_D :=  1/[2^{D+1}\pi^{D/2}\Gamma(D/2)]$ and primes denote derivatives with respect to $\hat\rho$.
Also, we get two equations for the anomalous dimensions, $\eta_\phi=\frac{16v_D\hat g^2}{D}N m_{4}^\mathrm F(2\hat g^2\hat\rho)$ and $\eta_\psi=\frac{8v_D\hat g^2}{D}m_{12}^\mathrm F(2\hat g^2\rho,u^\prime)$.
The threshold functions $l^\mathrm B$, $l^\mathrm F$, $m_4^\mathrm F$, $m_{12}^\mathrm F$, and $l_{11}^\mathrm{FB}$ depend on the choice of regulator~\cite{Litim:2000ci,Litim:2001up,Litim:2001fd,Litim:2002cf,PAWLOWSKI20072831,PAWLOWSKI2017165}.
We use linear cutoffs $R_\phi=Z_\phi p^2r_\phi$ and $R_\psi=Z_\psi \slashed{p} r_\psi$ of the form $r_{\psi}(q)=\big(\frac{k}{q}-1\big)\,\Theta(k^2-q^2)$ and \mbox{$r_{\phi}(q)=\big(\frac{k^2}{q^2}-1\big)\,\Theta(k^2-q^2)$}, where
$\Theta(x)$ is the step function.
Explicitly, the needed threshold functions for $u$ then read~\cite{Berges:2000ew}
\begin{align*}
l^{\mathrm B}(\omega)&=\frac{2}{D}\left(1-\frac{\eta_\phi}{D+2}\right)\frac{1}{1+\omega}\,,\\
l^{\mathrm F}(\omega)&=\frac{2}{D}\left(1-\frac{\eta_\psi}{D+1}\right)\frac{1}{1+\omega}\,.
\end{align*}
The threshold function for $\hat g^2$ is given by
\begin{multline*}
l_{11}^\mathrm{FB}(\omega_\psi,\omega_\phi) =
\frac{2}{D}\bigg[\left(1-\frac{\eta_\psi}{D+1}\right)\frac{1}{1+\omega_\psi} \\
+\left(1-\frac{\eta_\phi}{D+2}\right)\frac{1}{1+\omega_\phi}\bigg]\frac{1}{(1+\omega_\psi)(1+\omega_\phi)}\,,
\end{multline*}
and for the anomalous dimensions we get
\begin{align*}
m_4^{\mathrm F}(\omega)&=\frac{1}{(1+\omega)^4}+\frac{1-\eta_\psi}{D-2}\frac{1}{(1+\omega)^3}\notag\\
&\quad-\left(\frac{1-\eta_\psi}{2D-4}+\frac{1}{4}\right)\frac{1}{(1+\omega)^2}\,,\displaybreak[0]\\
m_{12}^{\mathrm{FB}}(\omega_\psi,\omega_\phi)&= \left(1-\frac{\eta_\phi}{D+1}\right) \frac{1}{(1+\omega_\psi)(1+\omega_\phi)^2}.
\end{align*}
Further, we expand the potential $u$ in powers of $\hat\rho$ around the origin,
$
u(\hat\rho)=\sum_{i=1}^{i_\text{max}}\frac{\hat\lambda_{i}}{i!} \hat\rho^i\,
$
which defines the couplings $\hat\lambda_i$.
The flow equations for the $\hat\lambda_{i}$ are obtained by the rule
%
$
	\partial_t \hat\lambda_{i} = \big(\frac{\partial^{i}}{\partial \hat\rho^i}\partial_t u(\hat\rho)\big)\big\vert_{\hat\rho=0}\,.
$
%
Similarly, $\partial_t \hat g^2$, $\eta_\phi$, and $\eta_\psi$ are obtained.
For the explicit values given in Tab.~\ref{tab:N20}, we have set $D=3, N=20$ and we have chosen $i_\mathrm{max}=6$.
We have checked that this provides converged fixed-point properties within the polynomial expansion of the potential.

\bibliography{interpolation}

\end{document}